\documentclass[sigconf]{acmart}
\usepackage{MnSymbol}
\usepackage[linesnumbered,ruled,vlined,algo2e]{algorithm2e}
\usepackage{algpseudocode}
\usepackage{algorithm}
\usepackage{textgreek}
\usepackage{tablefootnote}
\usepackage{multirow}
\usepackage{overpic}
\usepackage{subfigure}
\usepackage{enumitem}
\usepackage{bm}
\usepackage{xcolor}
\usepackage{tabu}
\usepackage{amsmath}
\usepackage{tablefootnote}
\usepackage{subcaption}
\usepackage{subfigure}
\usepackage{tcolorbox}
\usepackage{caption}
\usepackage[normalem]{ulem}

\newcommand{\etal}{\textit{et al.}\xspace}

\newcommand{\ie}{\textit{i.e.}\xspace}

\definecolor{ao(english)}{rgb}{0.0, 0.5, 0.0}
\newcounter{dingheng}
\numberwithin{dingheng}{section}

\newcounter{fan}
\numberwithin{fan}{section}

\newcounter{junfeng}
\numberwithin{junfeng}{section}

    \newcounter{Siqiang}
\numberwithin{Siqiang}{section}

\setcopyright{acmlicensed}
\copyrightyear{2025}
\acmYear{2025}
\acmDOI{XXXXXXX.XXXXXXX}

\acmConference[Conference acronym 'XX]{Make sure to enter the correct
  conference title from your rights confirmation emai}{June 03--05,
  2018}{Woodstock, NY}
\acmISBN{978-1-4503-XXXX-X/18/06}

\begin{document}

\title{[Experiments \& Analysis]\\ Evaluating Learned Indexes in LSM-tree Systems: Benchmarks, Insights and Design Choices}

\author{Junfeng Liu}
\affiliation{%
  \institution{Nanyang Technological University}
   \country{Singapore}
  }
\email{junfeng001@e.ntu.edu.sg}

\author{Jiarui Ye}
\affiliation{%
  \institution{Nanyang Technological University}
   \country{Singapore}
  }
\email{jiarui005@e.ntu.edu.sg}

\author{Mengshi Chen}
\affiliation{%
  \institution{Nanyang Technological University}
  \country{Singapore}
  }
\email{mengshi002@e.ntu.edu.sg}

\author{Meng Li}
\affiliation{%
  \institution{Nanjing University}
   \country{China}
  }
\email{meng@nju.edu.cn}

\author{Siqiang Luo}
\authornote{Corresponding Author}
\affiliation{%
  \institution{Nanyang Technological University}
   \country{Singapore}
  }
\email{siqiang.luo@ntu.edu.sg}

\begin{abstract}
LSM-tree-based data stores are widely used in industry due to their exceptional performance. 
However, as data volumes grow, efficiently querying large-scale databases becomes increasingly challenging. To address this, recent studies attempted to integrate learned indexes into LSM-trees to enhance lookup performance, which has demonstrated promising improvements. Despite this, only a limited range of learned index types has been considered, and the strengths and weaknesses of different learned indexes remain unclear, making them difficult for practical use. To fill this gap, we provide a comprehensive and systematic benchmark to pursue an in-depth understanding of learned indexes in LSM-tree systems. 
In this work, we summarize the workflow of 8 existing learned indexes and analyze the associated theoretical cost. We also identify several key factors that significantly influence the performance of learned indexes and conclude them with a novel configuration space, including various index types, boundary positions, and granularity. Moreover, we implement different learned index designs on a unified platform to evaluate across various configurations. Surprisingly, our experiments reveal several unexpected insights, such as the marginal lookup enhancement when allocating a large memory budget to learned indexes and modest retraining overhead of learned indexes. Besides, we also offer practical guidelines to help developers intelligently select and tune learned indexes for custom use cases.
\end{abstract}

\maketitle

\section{Introduction}
\label{sec:intro}
Log-structured Merge Trees (LSM-trees), have already been widely used as the fundamental storage structure underpinning many key-value stores like Google Spanner~\cite{corbett2013spanner}, Apache Cassandra~\cite{lakshman2010cassandra}, and MongoDB WiredTiger~\cite{wiredtiger}. These key-value stores play pivotal roles in various applications in social media, streaming processing, and file systems.

\begin{figure*}
    \centering
    \includegraphics[width=\linewidth]{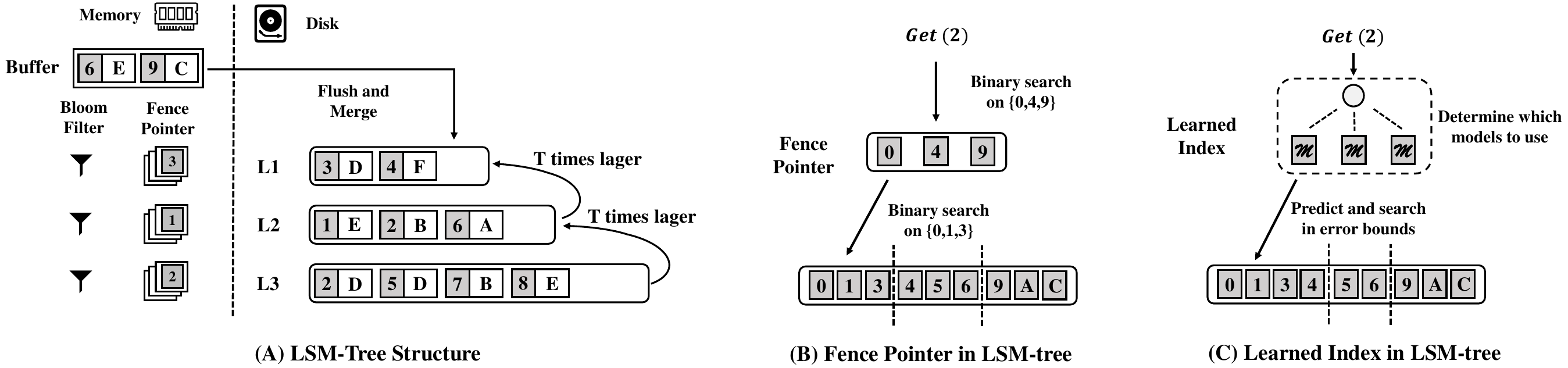}
    \caption{(A) presents the general structure of an LSM-tree; (B) and (C) illustrate how original fence pointer and learned indexes work in an LSM-tree, respectively}
    \label{fig:bg}
\end{figure*}

\noindent{\bf Index in LSM-tree stores.} As illustrated in Figure ~\ref{fig:bg}(A), a typical LSM-tree consists of both memory and disk components. The memory components include a write buffer, bloom filters, and fence pointers, while the disk components are sorted key-value arrays organized across multiple levels. To prevent the need for sequential scanning of these arrays on disk, the {\it fence pointers} in memory help quickly locate the specific data block where the target key is stored. This allows only the relevant data block to be read, thereby significantly reducing the number of I/Os required.

While fence pointers have successfully reduced the number of I/Os, the question still arises: can we push the performance further within the same memory constraints? Recent advances in machine learning have led to the development of several learned indexes, which have demonstrated superior memory efficiency through extensive evaluations~\cite{learnedindexbenchmark}. These learned indexes achieve impressive results using much less memory space, presenting an exciting opportunity to further optimize indexing in LSM-trees. However, two main questions remain unanswered when applying them to LSM-tree systems:

\noindent{\bf Are all learned indexes suitable for LSM-tree systems?}
Recent years have seen a surge of interest in integrating learned indexes into LSM-tree systems. Dai \etal~\cite{dai2020wisckey} are among the first to explore this integration by applying a Piece-wise Linear Regression (PLR) model to LevelDB~\cite{leveldb}, achieving significant improvements over traditional fence pointers. Similarly, Lu \etal~\cite{lu2021tridentkv} demonstrate promising results using the Recursive Model Index (RMI) within an LSM-tree framework. While these studies show the potential of learned indexes in LSM-tree systems, they also prompt a natural question: are all recently proposed learned indexes well-suited for use in LSM-tree architectures?
Evaluating this is nontrivial. Given the large and growing number of proposed learned index designs, implementing all of them within a single system for empirical comparison is infeasible. A more practical approach is to first identify the distinctive storage characteristics of LSM-tree systems, such as the hierarchical level and immutable files, then classify learned indexes based on their compatibility with these characteristics. This allows us to systematically determine which types of learned indexes are most appropriate for LSM-tree environments.

\noindent{\bf Is there a universal guideline and configuration space for tuning learned indexes in LSM-tree systems?}
While prior benchmarks~\cite{learnedindexbenchmark,kraska2018learnedIndex,updatableindexready} have outlined practical principles for deploying learned indexes in both in-memory and on-disk systems, tuning them within LSM-tree systems remains significantly more complex. This complexity arises from the presence of multiple interdependent components, such as Bloom filters, caches, and write buffers, which all compete for the same limited memory budget and jointly influence both update and lookup performance. How to effectively allocate memory between learned indexes and these LSM-specific components is still an open question.
Furthermore, the configuration space of learned indexes is highly diverse, with different models requiring different parameters and optimization strategies. While developing a universal tuning guideline would be highly valuable, it is also inherently challenging due to the variability in index structures and the interplay with system-level tradeoffs.

\vspace{1mm}

Motivated by these questions, in this paper, we conduct a comprehensive study on applying learned indexes in LSM-tree systems. Our contributions are concluded as the following:

\noindent{\bf We revisit several prevalent learned indexes and identify the key factors that impact the compatibility between learned indexes and LSM-tree.} We start with revisiting the existing learned indexes, summarizing their structures and underlying algorithms. Next, we classify these indexes based on their data layout to determine which ones are most compatible with LSM-tree architectures. Particularly, some representative learned indexes, such as ALEX~\cite{ding2020alex} and LIPP~\cite{lipp}, are primarily designed for in-memory lookups and involve less efficient pointer jumping, making their data layout incompatible with the continuous and sorted structure of LSM-tree levels. A detailed categorization and compatibility study of learned indexes tailored for LSM-trees would offer valuable insights for designing more suitable models for LSM-tree systems.

\noindent{\bf We identify three unified key parameters that affect performance across all compatible indexes and propose a comprehensive configuration space for LSM-trees.} Our theoretical analysis reveals three core tuning parameters that significantly influence the performance of learned indexes in LSM-tree systems: index type, position boundary, and index granularity.
Index type refers to the specific learned index model employed, each characterized by unique segment partitioning strategies and inner index structures, leading to different memory-performance tradeoffs.
Position boundary denotes the final search range that the LSM-tree retrieves from disk. This parameter directly affects I/O cost and is a crucial tuning knob for many learned index designs.
Index granularity determines the number of entries over which a learned index is constructed, influencing both lookup accuracy and index overhead.
These three parameters define a unified and inclusive configuration space that enables systematic experimentation and performance evaluation. This framework provides a solid foundation for understanding and optimizing the integration of learned indexes in LSM-tree systems.

\noindent{\bf We develop a unified testbed LSM-tree system with a learned-index-compatible interface that enables seamless integration and fair comparison of six representative indexes.}
In Section~\ref{sec:config}, we detail the implementation of a universal interface that allows diverse learned indexes to be easily integrated into LevelDB. We also illustrate how the three key parameters—index type, position boundary, and index granularity—affect LSM-tree performance. Using this platform, we conduct a comprehensive evaluation of six representative learned indexes compatible with LSM-trees, testing them under various configurations and datasets to assess their impact on core system operations such as point lookups, range lookups, and compaction.
Our findings yield several important insights. First, all evaluated learned indexes exhibit a superior memory-latency tradeoff compared to traditional fence pointers, offering significantly lower lookup latency for the same memory usage. This underscores the potential of learned indexes in LSM-tree systems. Among the tested models, RMI, and PGM consistently demonstrate strong performance and dominate other approaches in most scenarios.
We also observe that the overhead introduced by model training and write-time indexing during compaction is modest relative to the overall compaction cost. Additionally, increasing index granularity reduces the number of learned indexes required, thereby lowering overall memory consumption.
Beyond point queries, we further evaluate the effects of learned indexes on range lookups and mixed workloads involving both reads and updates. Based on these experiments, we identify three key design principles for integrating learned indexes into LSM-trees:
\begin{itemize}[leftmargin=*]
    \item Position boundary is the most critical factor for lookup performance. While the choice of index type mainly affects the memory-latency tradeoff, prioritizing models with smaller error bounds under a fixed memory budget provides greater benefits than optimizing internal index structures.
    \item Increasing SSTable size improves lookup performance under a fixed memory budget by reducing index memory overhead and enabling the use of smaller position boundaries.
    \item Memory allocation exhibits diminishing returns: the performance improvement from increasing the memory budget plateaus once the segment size becomes smaller than or equal to the I/O block size.
    \item The construction and training of learned indexes introduce minimal overhead compared to traditional fence pointers during compaction.
\end{itemize}
\section{Background}
\label{sec:background}

This section discusses the background knowledge about the LSM-tree systems and the indexing schemes over it.
\subsection{Log-Structured Merge Trees}
An LSM-tree efficiently manages data across multiple disk components, organizing data into sorted arrays at different levels. It also maintains an in-memory component, where recent updates are stored in a write buffer. To enhance lookup performance, each sorted array is associated with Bloom filters and indexes. The capacity of each level increases exponentially by a size ratio of $T$, meaning the total number of levels required to store $N$ entries is approximately $L = \lceil \log_{T} \frac{N \cdot e}{F} \rceil$, where $F$ is the size of the write buffer, and $e$ is the size of individual entries. Each level consists of key-value pairs stored in a sorted array, referred to as a {\it sorted run} in some works. LSM-trees primarily support three operations:

\noindent{\bf Updates.} Updates in an LSM-tree are initially written to the in-memory write buffer. Once the buffer is full, the key-value entries are flushed to disk and merged into the sorted array at level-1, as shown in Figure ~\ref{fig:bg} (A). If a level exceeds its capacity, a compaction operation is triggered, merging entries into the next level to maintain efficient storage and query performance. Alternatively, to mitigate resource consumption spikes, some databases~\cite{leveldb,rocksdb} perform partial compactions by merging only a subset of entries into the next level, rather than compacting the entire level at once. In these systems, a sorted run is divided into sorted files, known as {\it SSTables}, and only a subset of these SSTables is selected for merging during each compaction.

\noindent{\bf Point Lookups.} A point lookup searches for the value of a specific key by checking levels sequentially until the key is found. To expedite this process, Bloom filters and indexes are employed. When a Bloom filter indicates a potential match, the LSM-tree locates the approximate range within that level and performs a binary search to verify the key’s presence.

\noindent{\bf Range Lookups.} 
Range lookups collect entries from each LSM-tree level and use a sort-merge process to remove duplicates, returning only the most recent values.

\subsection{Indexes in LSM-tree}
Traditional index structures in LSM-trees often rely on {\it fence pointers}, as shown in Figure ~\ref{fig:bg}(B). These pointers store the smallest (or largest) key of a fixed range of key-value pairs, allowing the LSM-tree to quickly narrow down the search to a small data range when locating a specific key. During compaction, the smallest or largest key of each newly created data range is stored in memory. By querying these index structures, the LSM-tree can skip unnecessary searches through large amounts of data on disk, significantly reducing I/O operations and improving lookup efficiency.

\noindent{\bf Learned Indexes in LSM-tree.} Although fence pointers are widely used in most LSM-tree systems~\cite{leveldb,rocksdb,lakshman2010cassandra}, there remains an opportunity to improve memory efficiency by replacing them with more advanced learned index structures. This potential arises for two key reasons: (1) The sorted arrays on disk are immutable, meaning they are only created and deleted during compactions, making them well-suited for even non-updatable learned indexes, and (2) since the entries are already stored in a naturally sorted order on disk, learned indexes can efficiently map the data, potentially reducing the overhead of sorting. As shown in Figure~\ref{fig:bg} (C), by training the learned model during compactions, we can easily replace the fence pointers with learned indexes.

Abu-Libdeh \etal~\cite{googlelearned} were the first to evaluate the feasibility of using a linear regression model in LSM-tree systems, finding a positive impact when replacing traditional fence pointers with learned index structures. However, their study did not systematically explore the performance variations across different types of learned indexes or how various configurations might affect results. Building on this idea, Dai \etal integrated learned indexes into a key-value separated LSM-tree system, Wisckey~\cite{lu2017wisckey}, developing Bourbon~\cite{dai2020wisckey}, an LSM-tree system equipped with a piecewise linear learned index. While Bourbon demonstrated notable performance improvements, it still did not thoroughly investigate all possible design choices for learned indexes, such as experimenting with different index types.
\section{Learned Indexes Revisited}
\begin{figure*}
    \vspace{-2mm}
    \centering
    \includegraphics[width=\linewidth]{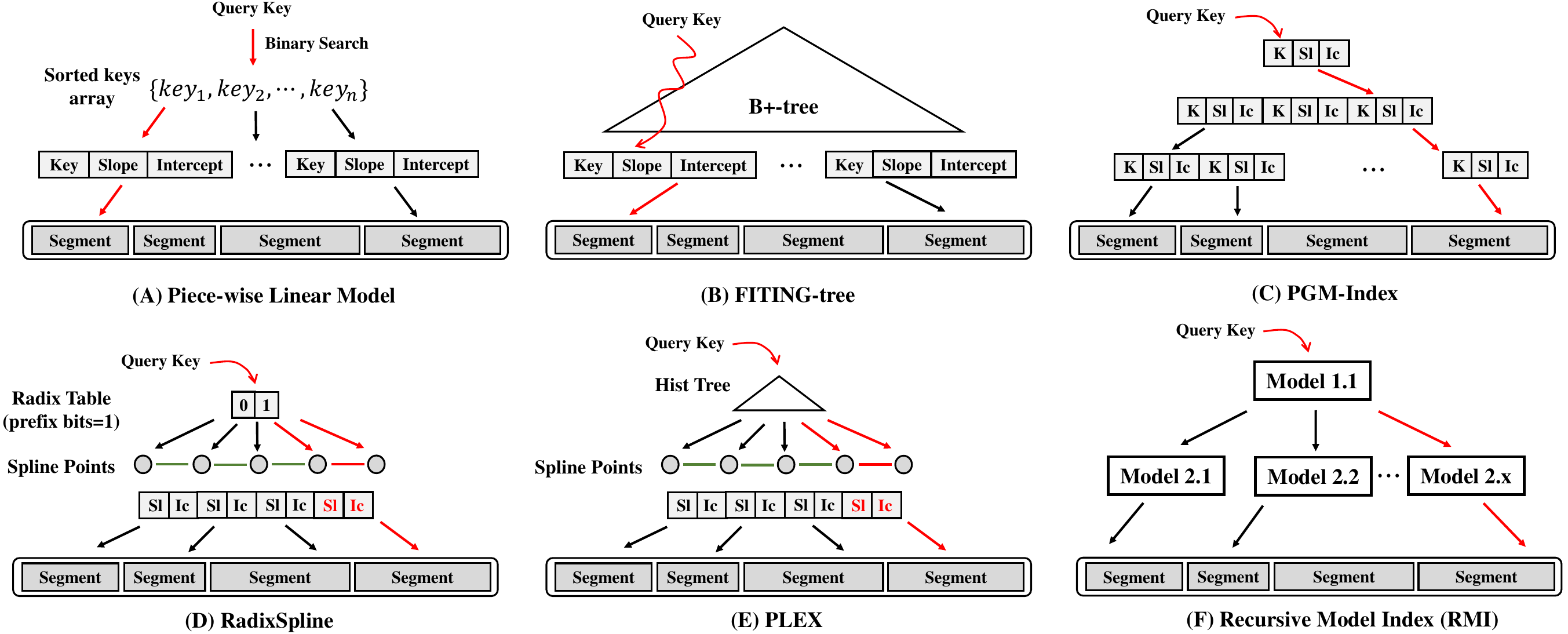}
     \vspace{-2mm}
    \caption{(A) to (F) present the data structures of data-clustered learned indexes and the general lookup procedure. {\bf Ic} represents the intercept of the linear model, {\bf Sl} is the slope, and {\bf K} is short for the key.}
    \label{fig:learnedindexintro}
     \vspace{-2mm}
\end{figure*}
\label{sec:revisit}
We revisit eight learned indexes in detail and assess their compatibility with LSM-tree systems. Broadly, we classify these indexes into two categories based on their data layout: {\bf data-clustered indexes}, such as FITing-Tree, PGM, and RMI, and {\bf data-unclustered indexes}, including LIPP and ALEX. As illustrated in Figure ~\ref{fig:learnedindexintro}, data-clustered indexes store key-value pairs in physically continuous blocks, whereas, as shown in Figure ~\ref{fig:unclustered}, data-unclustered indexes do not. Instead, data-unclustered indexes require additional steps, such as traversing via pointers, to retrieve continuous key-value pairs. In the following sections, we first review the structures of the most representative data-clustered and data-unclustered indexes. We then analyze which indexes are most compatible with LSM-tree systems, followed by a theoretical analysis of the cost associated with each index.

\subsection{Data Clustered Indexes}
In this subsection, we review six well-known data-clustered learned indexes and their respective lookup procedures.

\noindent{\bf Piece-wise Linear Regression (PLR) ~\cite{dai2020wisckey}}, shown in Figure ~\ref{fig:learnedindexintro} (A), uses a greedy algorithm to divide a sorted array into segments based on a specified error bound, which represents the maximum allowable difference between the estimated and actual key positions. For each segment, PLR builds a linear model to predict the approximate position of a key. During a lookup, PLR first locates the segment containing the key by performing a binary search over the segments. It then uses the corresponding linear model to estimate the key's position, denoted as {\it appx\_pos}. Since the linear model ensures that the true key position lies within a range of [{\it appx\_pos - error}, {\it appx\_pos + error}]--where the {\it error} reflects the prediction tolerance--a binary search is performed within this range to find the exact key location.

\noindent{\bf FITing-Tree ~\cite{fitingtree}}, shown in Figure ~\ref{fig:learnedindexintro} (B), uses a greedy algorithm to divide a sorted array into segments similarly based on a specified error bound. A linear model is built for each segment to predict the position of keys. To efficiently search through these segments, FITing-Tree uses a B+-tree to index the segments. When looking up a key, FITING-tree first traverses the B+-tree to locate the segment containing the key. It then uses the corresponding linear model to predict the approximate position. Since the linear model ensures that the true key lies within the range [{\it appx\_pos - error}, {\it appx\_pos + error}], a binary search is performed within this range to find the exact key location.

\noindent{\bf Piecewise Geometric Model index (PGM)~\cite{ferragina2020pgm}}, shown in Figure ~\ref{fig:learnedindexintro} (C), takes a different approach by using a streaming algorithm, rather than a greedy one, to divide the array into segments and build linear models with a given error bound. PGM further applies this streaming algorithm recursively to construct parent nodes and build linear models for these higher-level nodes. To look up a key, PGM predicts an approximate position {\it appx\_pos} using its model, then recursively performs a binary search within [{\it appx\_pos - error}, {\it appx\_pos + error}] until the exact key is found in the leaf node.

\noindent{\bf RadixSpline (RS)~\cite{kipf2020radixspline}}, shown in Figure ~\ref{fig:learnedindexintro} (D), selects a subset of keys from the sorted array as spline points and uses linear interpolation models to estimate the positions of keys between any two spline points. RadixSpline ensures the accuracy of its spline layer by imposing error bounds on the approximations. If the error exceeds a predefined threshold, additional spline points are added to improve the approximation. To index these spline points, RadixSpline constructs a radix table. When looking up a key, RadixSpline first uses the radix table to locate the correct spline segment, then applies the linear interpolation model to predict the key’s approximate position. A binary search is then performed within the range [{\it appx\_pos - error}, {\it appx\_pos + error}] to locate the exact key.

\noindent{\bf Practical Learned Index (PLEX)~\cite{plex}}, shown in Figure ~\ref{fig:learnedindexintro} (E), is an improved version of RadixSpline that employs a hierarchical Hist Tree (or Radix Tree) to index spline points and reduce search space. Like RadixSpline, PLEX uses spline points and linear interpolation models to estimate key positions but improves lookup efficiency by leveraging this hierarchical structure. A key feature of PLEX is its self-tuning capability, which dynamically adjusts the number of spline points based on data distribution and workload. This optimization ensures a balance between prediction accuracy and memory usage, improving overall performance. During lookup, PLEX first uses the Hist Tree to locate the segment containing the key, then applies the corresponding linear model to predict the approximate position. A binary search is then performed within [{\it appx\_pos - error}, {\it appx\_pos + error}] to find the exact key. With its hierarchical structure and self-tuning, PLEX efficiently adapts to large datasets and varying workloads, offering better scalability than RadixSpline.

\noindent{\bf Recursive Model Index (RMI)~\cite{rmi}}, shown in Figure ~\ref{fig:learnedindexintro} (F), is a learned index that recursively applies machine learning models to approximate the position of keys in a sorted array. It organizes models in a hierarchical structure, where upper-level models predict the position of keys for the next layer, progressively refining the prediction until the lowest layer estimates the key’s position. RMI is built in a top-down manner, where the top-level model is trained first to give a coarse estimate of key positions. Based on this, the dataset is divided, and lower-level models are trained on smaller subsets, improving accuracy as you move down the hierarchy. This approach allows RMI to tailor the complexity of each model to the portion of the data it handles, optimizing both performance and memory usage. During lookup, RMI first uses the top-level model to make a rough prediction, then refines this through subsequent layers. The final model predicts the approximate key position, followed by a binary search within a small range to find the exact key. RMI’s error is not predefined by the user but rather recorded during the training process, adapting to the data’s characteristics.

\subsection{Data Unclustered Indexes}
\begin{figure}
    \centering
    \includegraphics[width=\linewidth]{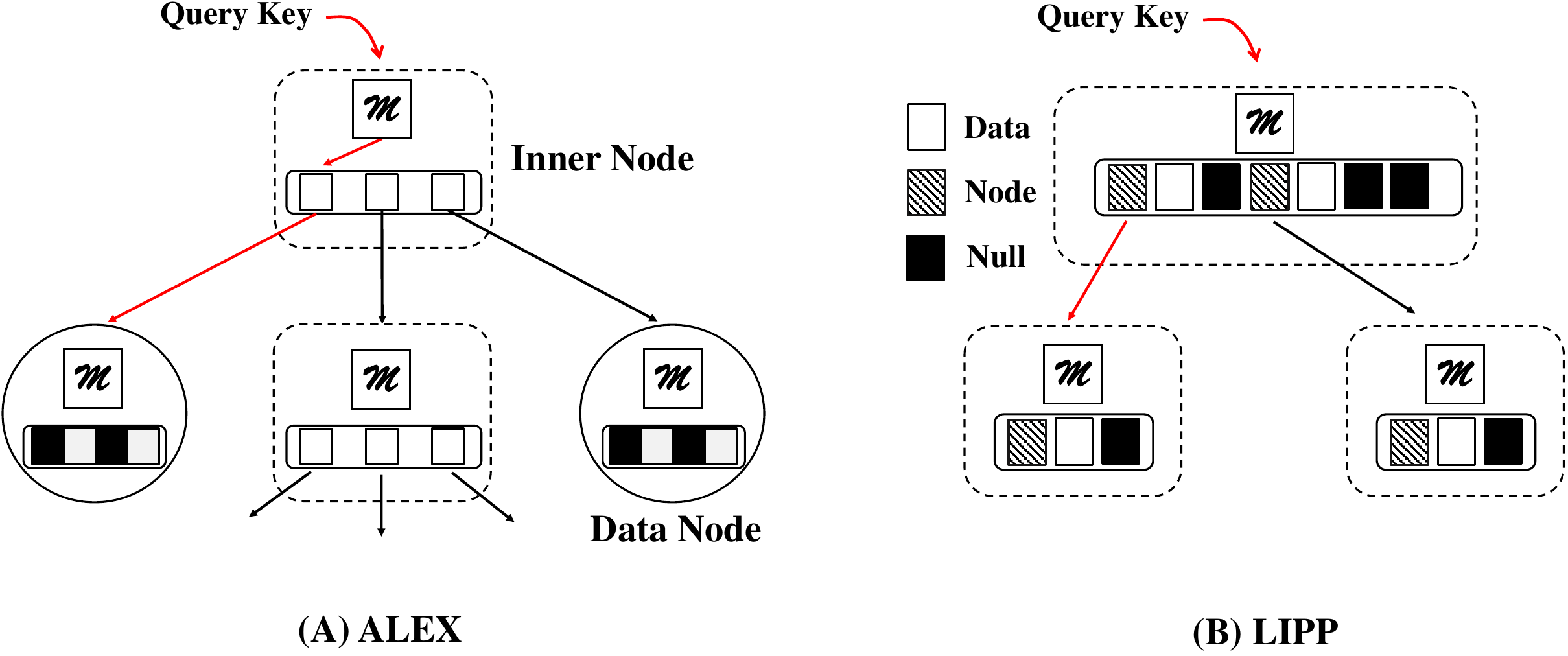}
    \caption{(A) and (B) present the structure of ALEX and LIPP. The data (key-value pairs) are not stored continuously to accommodate incoming new entries.}
    \label{fig:unclustered}
\end{figure}
\noindent{\bf ALEX~\cite{ding2020alex}}, shown in Figure ~\ref{fig:unclustered} (A), uses two types of nodes: inner nodes and data nodes, both combining arrays with linear models to predict key positions. Inner nodes contain an array of pointers to other nodes, while data nodes use gapped arrays that store key-value pairs, interleaving empty slots to support efficient insertions. During a lookup, ALEX traverses from the root through inner nodes, using the model to predict the appropriate child node. Once at a data node, the model predicts the position of the key, followed by an exponential search to locate the exact key. ALEX dynamically adjusts its structure by splitting or merging nodes as data evolves, ensuring efficient handling of both lookups and updates while maintaining optimal performance for dynamic workloads.

\noindent{\bf LIPP~\cite{lipp}}, shown in Figure ~\ref{fig:unclustered} (B), uses a linear model in each node to precisely predict key positions. Each node contains a data array and a bitmap to distinguish between three slot types: DATA, NULL, and NODE. The linear model predicts which slot should be accessed during a lookup. LIPP employs the Fastest Minimum Conflict Degree (FMCD) algorithm to minimize conflicts (multiple keys mapped to the same slot). When a key is inserted into a NULL slot, it is marked as DATA and stores the key-value pair. If inserted into an occupied DATA slot, the slot becomes a NODE, pointing to a new child node created with the conflicting keys. The child node is built using the same algorithm. During lookup, LIPP uses the root node’s linear model to predict the key’s position. If the predicted slot is NODE, it follows the pointer to the child node and repeats the process. If the slot is DATA, it checks for a match and returns the key-value pair if found.

\noindent{\bf DILI~\cite{li2023dili}} uses a two-phase approach to build the index: a bottom-up tree-building process using linear regression models based on global and local key distributions, followed by a top-down refinement where the fanout of internal nodes is customized according to local key distributions. This design strikes a balance between the number of leaf nodes and the tree height, both crucial factors in minimizing key search time. Additionally, DILI includes flexible algorithms for efficient key insertion and deletion, allowing the index to dynamically adjust its structure when necessary.

\noindent{\bf NFL~\cite{wu2022nfl}} introduces a new approach to addressing the challenges of learned indexes by transforming complex key distributions before constructing the index. NFL uses a two-stage framework: first, it applies Numerical Normalizing Flow (Numerical NF) to transform the key distribution into a near-uniform one. Then, it builds a learned index using a specialized After-Flow Learned Index (AFLI), optimized for the normalized data.

\subsection{LSM-Compatible Indexes}

Though data-unclustered indexes offer excellent read performance and memory efficiency, we argue that data-clustered indexes are a more suitable choice for LSM-trees. This is because LSM-trees rely on the sorted and contiguous storage of data across levels (i.e., SSTables and sorted runs) to optimize read and write performance. Data-clustered indexes naturally maintain this physical continuity, allowing them to replace existing fence pointers with minimal engineering effort. By contrast, data-unclustered indexes like ALEX and LIPP disrupt the current data layout (i.e., SSTable), making integration with LSM-trees more complex. Moreover, range lookups, a critical operation in LSM-trees, benefit from the sequential access provided by data-clustered indexes. These indexes ensure fast access to key-value pairs stored contiguously, while data-unclustered indexes, which scatter data, require additional memory and disk jumps, significantly increasing the overhead of such queries.

While integrating data-unclustered indexes into LSM-trees is a promising and intriguing topic, this integration is not feasible without significantly altering the widely recognized LSM-tree storage architecture for the following reasons:
\begin{enumerate}
    \item Need for non-compact and uncontinous storage layout replacement: Successful integration would necessitate replacing the current compact data layout (\ie, SSTables) with a discontinous data-unclustered structures, which represents a considerable undertaking.
    \item Re-Implementation of Basic Functions: Basic functions, such as range lookup and compaction iterators, would need to be re-implemented when using data-unclustered indexes.
\end{enumerate}
\begin{figure*}
    \centering
    \includegraphics[width=\linewidth]{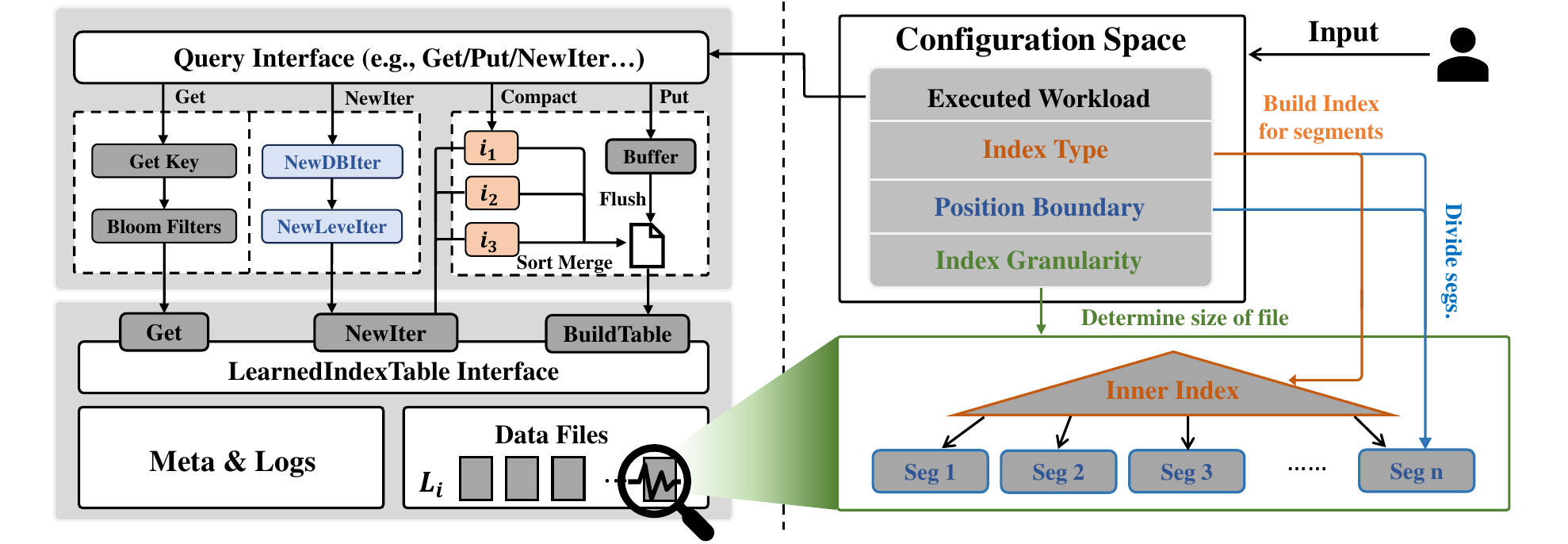}
    \caption{The figure demonstrates the architecture of our testbed system. The left-hand side is the detailed conduction of different operations to test while the right-hand side is how the three configuration impact the system.}
    \label{fig:pipeline}
\end{figure*}
Given these considerations, this paper primarily focuses on the performance of data-clustered indexes within existing LSM-tree systems, while leaving the exploration of data-unclustered indexes for future research.

\section{Exploring the Configuration Space of Indexing in LSM-trees}
\label{sec:config}
In this section, we are going to analyze the cost of LSM-compatible indexes to identify the tuning options that affect the read performance and memory consumption when applied to LSM-tree systems.

\subsection{Configuration Space} 
To begin with, data-clustered indexes locate a key by first identifying the segment likely to contain it, then reading that segment from disk, and finally searching within it under a bounded error range, as illustrated in Figure~\ref{fig:pipeline}. The cost of the first step depends on the type of index used, since different index structures organize segment metadata differently. The second cost, related to I/O, is primarily determined by the error bound. Data-clustered indexes ensure that segment lengths do not exceed $2\times$ the error bound ($\epsilon$), and thus, the I/O cost would not exceed $O(\frac{2\epsilon}{B})$, where $B$ is the size of an I/O block. The final cost stems from the in-segment search, typically performed using binary search, and is also governed by the error bound.
Based on this process, the two most critical factors to tune are:

\noindent{\bf Index Type.} The choice of index structure has a direct impact on lookup speed and memory overhead. For example, FITing-tree uses a B+-tree to index segments, which offers faster lookups but incurs higher memory consumption than simpler alternatives like the sorted arrays used in PLR. Thus, selecting the appropriate index type involves balancing performance and memory usage.

\noindent{\bf Position Boundary.} Once the approximate segment is identified, the LSM-tree must read it from disk. We define the ``position boundary'' as the range length of this segment, which is usually $2\times$ the error bound in the learned indexes. A smaller position boundary reduces I/O cost, which is often the dominant overhead in LSM-trees, but increases the number of segments, which raises memory usage. Tuning this boundary is therefore crucial for achieving an optimal tradeoff between I/O efficiency and memory consumption.

Beyond these two main factors, {\bf Index Granularity} also plays an important role in determining the performance of learned indexes within LSM-tree systems. Modern LSM-tree implementations (e.g., LevelDB, RocksDB, PebblesDB~\cite{leveldb,rocksdb,raju2017pebblesdb}) often apply partial compaction strategies, where sorted runs are divided into multiple files (SSTables), and only some are compacted into the next level. In such cases, learned indexes are typically built at the SSTable level, meaning their position boundary is bounded by the SSTable size. Dai \etal~\cite{dai2020wisckey} suggest that coarser-grained indexing—such as level-grained models like LevelModel—can yield performance improvements of around 10\% under read-heavy workloads. To examine this claim, we evaluate the performance of learned indexes built at varying SSTable sizes, as well as those constructed across entire levels instead of individual files.


\noindent{\bf Configuration Tradeoff.} Selecting an index type entails a tradeoff between memory consumption and lookup latency. To fairly evaluate index types, it is important to compare them under consistent configurations. Position boundary is likely the most influential parameter for read performance due to its impact on I/O cost. While reducing position boundaries can improve I/O efficiency, this often leads to increased memory usage by creating more, smaller segments. The memory overhead introduced by this tradeoff remains an open question and warrants empirical evaluation. Moreover, although increasing SSTable size (i.e., index granularity) reduces the number of indexes, it may also lead to less efficient compactions by increasing the volume of data processed per compaction round. Therefore, careful tuning of index granularity is also essential for optimizing read and compaction performance.

\subsection{Implementation}
To systematically evaluate the effectiveness of the three aforementioned configurations, we build a benchmark system based on LevelDB, a well-known and streamlined LSM-tree implementation. To integrate learned indexes without disrupting the system’s core functionality, we implement a new class, \texttt{LearnedIndexTable}, which inherits from and replaces the original \texttt{Table} class. We override three key functions: \texttt{InternalGet} (Get), \texttt{NewIterator} (NewIter), and \texttt{TableBuilder} (BuildTable). Below, we describe each implementation in detail and explain how these functions interact with the rest of the system.

\noindent{\bf Get.}
The \texttt{InternalGet} function handles point lookups by locating a specific key in a data file. In our implementation, key-value pairs are sorted and stored in segments, each indexed by a learned model. To perform a lookup, the learned index first consults its internal model to identify the corresponding segment. The segment is then fetched from disk using the Linux \texttt{pread} interface, and a binary search is conducted within the segment to retrieve the target key.

\noindent{\bf NewIter.}
The iterator interface is essential in LSM-tree systems, supporting both range queries and compaction. Our learned index iterator begins by seeking to a target key using the same procedure as \texttt{InternalGet}, and then proceeds to iterate over subsequent key-value pairs within the segment. Once all entries in the current segment are exhausted, the iterator advances to the next segment.

\noindent{\bf BuildTable.}
The \texttt{TableBuilder} interface is responsible for constructing learned-index-based tables. During flushes or compactions, the builder receives sorted key-value pairs and constructs a learned index over them, which is similar to how traditional fence pointers are built in baseline LevelDB. Additionally, the original on-disk SSTable format is replaced by the \texttt{LearnedIndexTable} format, in which the inner index and data segments are serialized separately, with their offsets recorded in the file header.

\begin{figure}
    \centering
    \includegraphics[width=\linewidth]{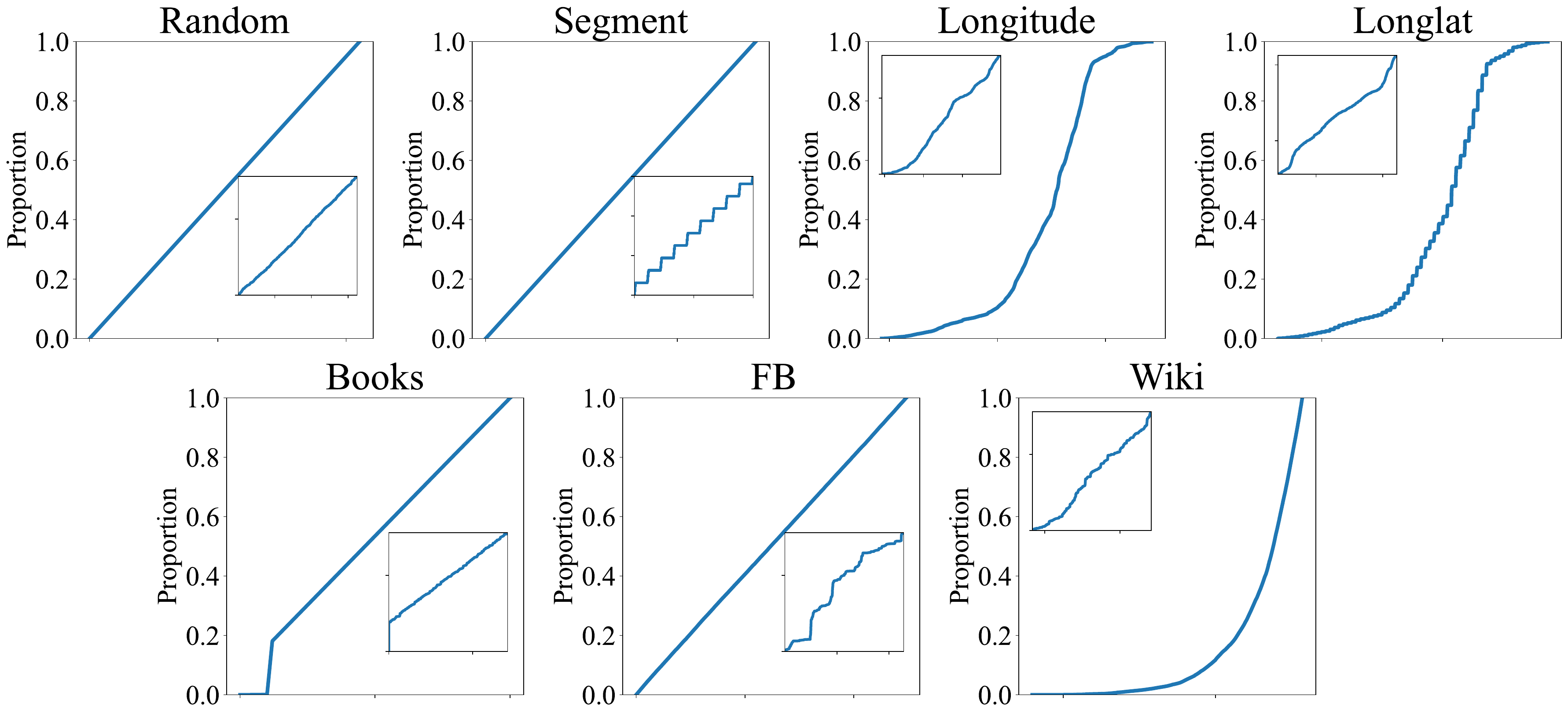}
    \caption{The CDF of different datasets}
    \label{fig:setup}
\end{figure}

Specifically, by using the interface, some commonly used LSM-tree processes and operations are implemented as the following:

\noindent{\bf Compaction.} The process of building learned indexes is similar to that of fence pointers: when an new table is created through a flush or compaction, the sorted key-value pairs are used to train the corresponding learned index. The index is then serialized, written to disk, and marked as ``learned''. Once the table is complete, the learned index is linked to it.

\noindent{\bf Point Lookup.} When a file is marked as ``learned'', LevelDB first locates the table containing the key and accesses the learned index to get a prediction for the target key. An I/O operation is then performed to fetch the approximate segment, followed by a binary search to retrieve the exact key.

\noindent{\bf Range Lookup.} A range lookup involves two phases: (1) identifying the starting point of the range at each level, and (2) retrieving the key-value pairs at each level and merging them until the entire range is fetched. The first phase is similar to a point lookup, involving access to the learned index and a binary search. The second phase differs slightly from LevelDB’s default behavior. We retrieve one I/O block at a time for the starting segment (typically 4096B) for all indexes until all key-value pairs in the target range are fetched from disk.

We integrate the following baselines by implementing the above interface into our system.
The implementations of PGM\footnote{https://github.com/gvinciguerra/PGM-index}, RadixSpline\footnote{https://github.com/learnedsystems/RadixSpline}, PLEX\footnote{https://github.com/stoianmihail/PLEX}, and PLR\footnote{We contacted the author to obtain the code}~\cite{dai2020wisckey} are based on versions released by the respective authors. For RMI and FITing-Tree, as no suitable C++ versions are available, we used the RMI implementation from a benchmark paper\footnote{https://github.com/BigDataAnalyticsGroup/analysis-rmi}~\cite{rmianalytics} and the FITing-Tree\footnote{https://github.com/RKolla99/FITing-Tree} implementation from the SOSD benchmark~\cite{sosd-vldb}.
\newcommand{\framedtext}[1]{%
\par%
\noindent\fbox{%
    \parbox{\dimexpr\linewidth-2\fboxsep-2\fboxrule}{#1}%
}%
}
\tcbset{colframe=black!75!black, boxrule=0.3mm, left=0.4mm, right=0.4mm, top=0.4mm, bottom=0.4mm, arc=0.75mm, after skip=3mm, before skip=2mm}

\section{Evaluation}
\begin{figure*}
    \centering
    \includegraphics[width=\linewidth]{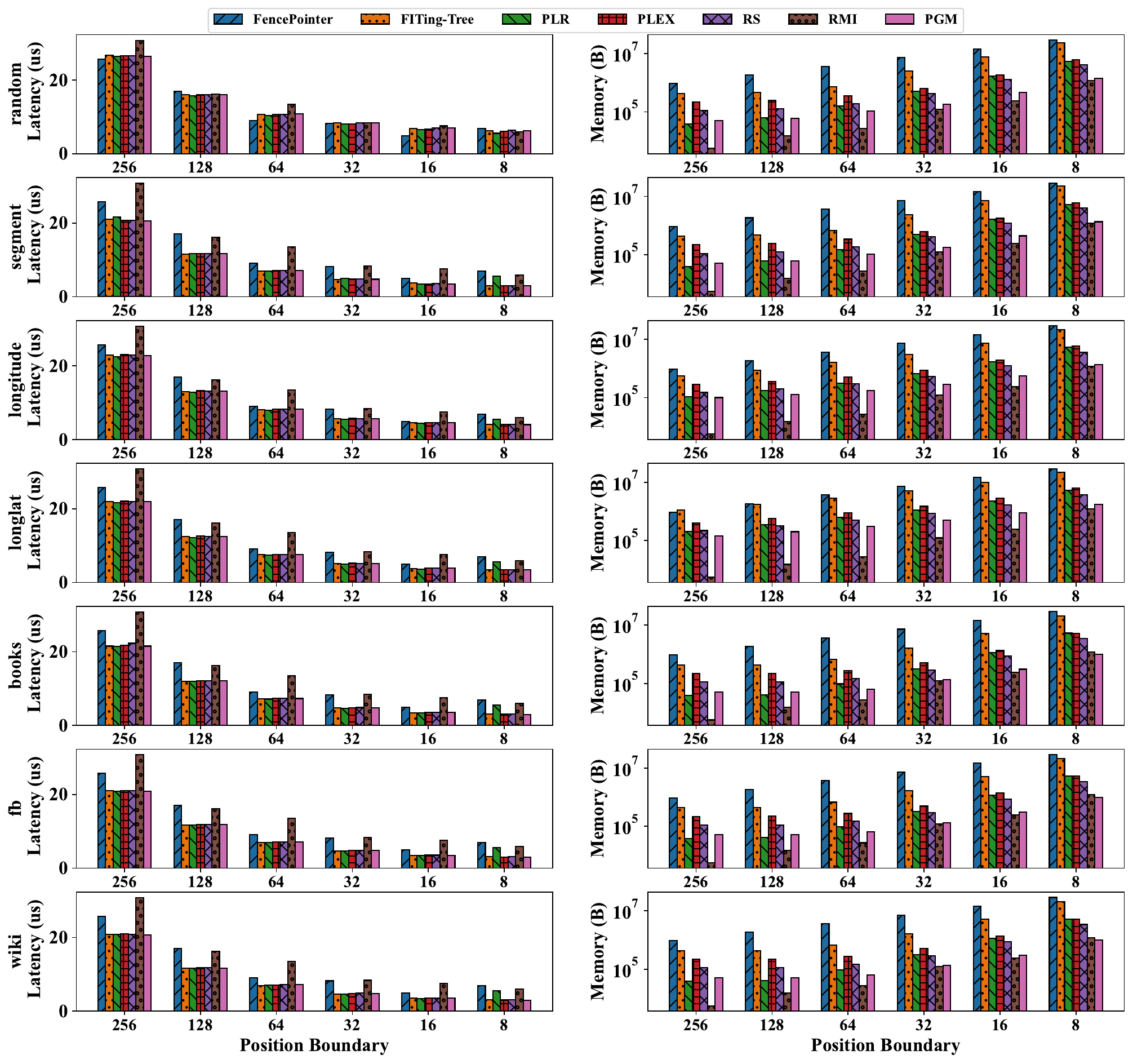}
    \caption{Latency and memory usage of different indexes under different position boundary under different datasets.}
    \label{fig:pointquery}
\end{figure*}
\begin{figure}
    \centering
    \includegraphics[width=\linewidth]{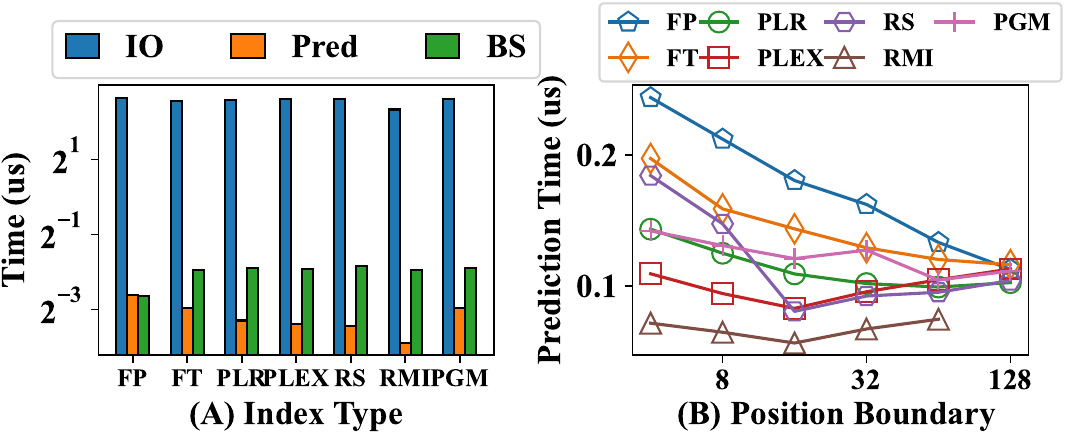}
    \caption{Query time breakdown}
    \label{fig:breakdown}
\end{figure}
\label{sec:evaluation}
\noindent{\bf Experiment Design.} In this section, we evaluate the performance of learned indexes across various scenarios. First, we assess the memory-latency tradeoff in lookups to determine the effectiveness of replacing fence pointers with learned indexes and to identify the key factors affecting lookup performance. Next, we address a common concern about the additional training time introduced during compaction by running a write-only workload and measuring compaction time. Following this, we evaluate unique features of LSM-tree systems, such as the asymmetric read overhead across different levels and the impact on range lookups. Finally, we test the performance under mixed workloads using the YCSB benchmark to assess their effectiveness in real-world applications.

\noindent\textbf{Datasets and Workload Setup} We evaluate all the indexes on seven different datasets generated by the SOSD benchmark~\cite{sosd-vldb}: Random, Segment, Longitude, Longlat, Books, FB, and Wiki with their cumulative distributions shown in Figure \ref{fig:setup} (a). Due to space constraints, we present results for the Random dataset in this paper. For a complete set of results on all datasets, please refer to our technical report~\cite{technicalreport}. Each dataset consists of 6.4 million key-value pairs, with 24-byte keys and 1000-byte values. In each experiment, we perform 1,000,000 operations.

\noindent\textbf{Running Environment.} 
We conduct our experiments on a machine running Ubuntu 22.04, equipped with an Intel Core i9-13900K CPU (36 MB L3 cache), 128 GB of memory, and a 2 TB NVMe SSD. All learned indexes are integrated into LevelDB, with the LSM-tree configured to use a leveling compaction policy, a size ratio of 10, and a 10-bit-per-key Bloom filter.

\noindent\textbf{Settings of Learned Indexes.} To evaluate the performance of fence pointers under different position boundaries, we adjust the data block size in LevelDB to generate varying numbers of fence pointers {\bf (abbr. FP)}. For PLR, FITing-Tree {\bf (abbr. FT)}, and PLEX, we directly vary the error bounds to control the position boundaries. For RMI, we follow the guidelines in ~\cite{rmianalytics} and use their {\it RMILabs} implementation, as recommended in the paper. This setup uses a two-level model tree. To vary the position boundary, we adjust the size of the second level, which in turn affects the position boundary. 
For both RadixSpline {\bf (abbr. RS)} and PGM, the error bounds can be adjusted to control the position boundaries. However, since both have additional parameters that affect their internal structure, these must also be fine-tuned for optimal performance. In PGM, the {\it EpsilonRecursive} parameter defines the error bound for internal nodes. We test various values, and find that {\it EpsilonRecursive} has little impact on PGM's performance in LSM-tree systems. Therefore, we retain the default setting of $EpsilonRecursive=4$. For RS, the {\it RadixBits} parameter controls the size of its radix table. After varying this value, we determine that $RadixBits=1$ offers the best tradeoff in LSM-tree systems, reducing memory usage while maintaining satisfactory performance.

\begin{figure*}
    \centering
    \includegraphics[width=\linewidth]{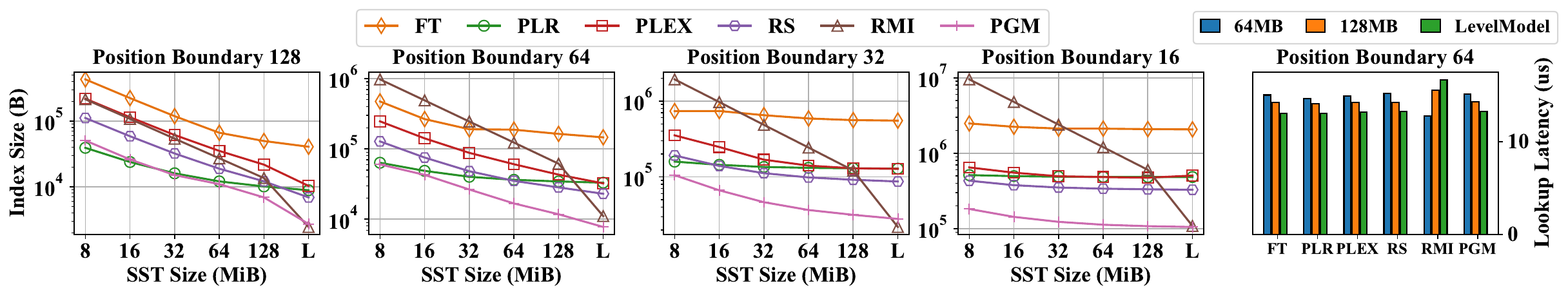}
    \caption{Impact of index granularity on point lookup.}
    \label{fig:granularity}
\end{figure*}
\begin{figure}
    \centering
    \includegraphics[width=\linewidth]{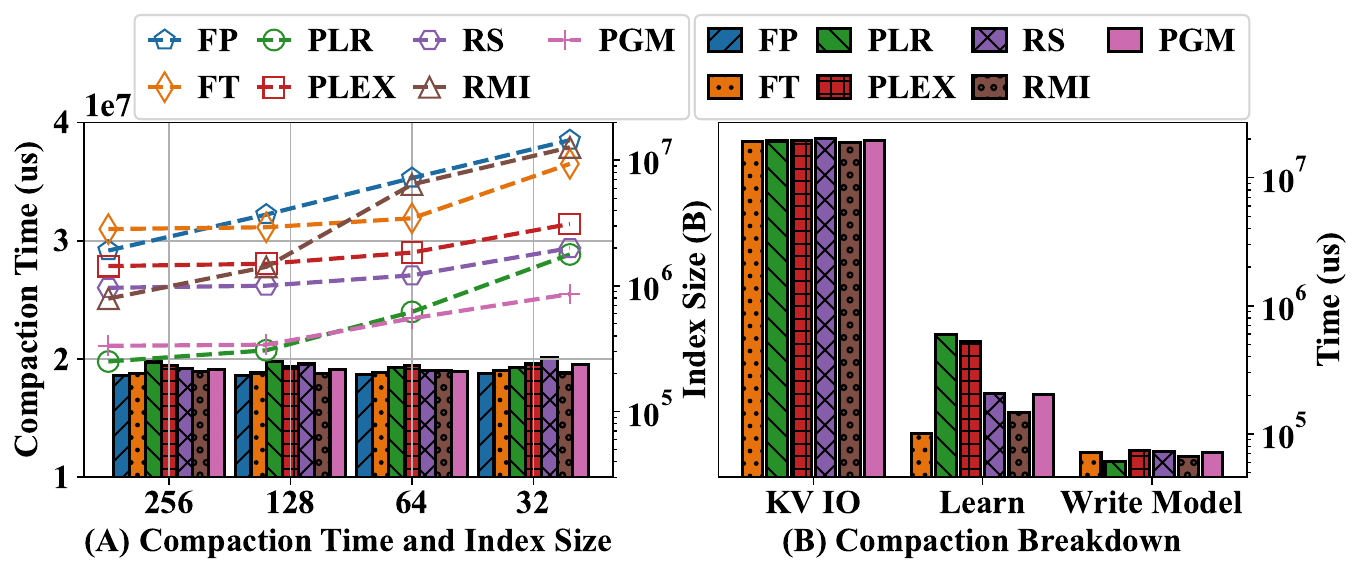}
    \caption{Compaction time and breakdown.}
    \label{fig:compaction}
\end{figure}
\subsection{Analysis on Position Boundary}\label{sec:pareto}

\begin{tcolorbox}
\noindent {\bf Observation 1}: Smaller position boundary positively reduces the latency for all indexes at the cost of increasing the memory usage. The efficiency of improving performance by adding memory budget varies from indexes, indicating the memory-latency tradeoff is very different.
\end{tcolorbox}
\vspace{-2mm}
\noindent{We vary the position boundary from 256 to 8 for each index and evaluate their performance under a point lookup workload consisting of 1,000,000 queries. We record both the lookup latency and memory usage, as shown in Figure~\ref{fig:pointquery} (A) and (B). Overall, decreasing the position boundary reduces lookup latency but increases memory consumption.

When the position boundary is held constant, all indexes exhibit nearly identical lookup latencies. This suggests that I/O dominates the cost of point lookups, while the time spent accessing the inner index and searching within a segment is relatively minor. As illustrated in Figure~\ref{fig:breakdown}(A) and (B), the I/O time required to fetch the segment from disk is approximately 10 times greater than the combined time for model prediction (including inner index access) and binary search within the segment. Although model prediction time increases slightly as the position boundary decreases (Figure~\ref{fig:breakdown}(B)), this increase is outweighed by the I/O reduction, resulting in an overall latency improvement for all indexes.

Despite similar lookup performance, memory usage varies significantly across index types. Traditional fence pointers exhibit the worst memory-latency tradeoff: as the position boundary decreases, their memory usage grows rapidly compared to learned indexes. This aligns with one of the core design goals of learned indexes—to reduce memory overhead.
However, the specific design of a learned index also impacts its memory-latency tradeoff. For instance, while FITing-Tree outperforms fence pointers, its memory consumption still increases quickly. This is due to its use of a B+-tree structure, which improves prediction accuracy at the cost of additional memory. RadixSpline (RS) and PLEX demonstrate comparable tradeoffs; both divide sorted data using spline points and use a radix table (RS) or radix tree (PLEX) to index those points. Although PLEX incorporates a self-tuning mechanism to optimize radix tree performance, this benefit is less pronounced in LSM-tree systems. This is likely because PLEX’s optimizations are most effective under skewed key distributions, which are uncommon in LSM-trees—especially at levels with fewer keys.
PLR shows a memory-latency tradeoff similar to RS and PLEX, largely due to its lightweight inner index structure for leaf segments, which helps conserve memory. Among all the evaluated learned indexes, PGM and RMI achieve the best memory-latency tradeoffs. PGM employs an optimized segmenting algorithm that reduces the number of segments needed for a given error bound, thus lowering memory usage. RMI stands out for its flexible configuration: by setting a large second-level index, it can achieve extremely low error bounds (as small as 1), enabling high precision with minimal memory. In contrast, other indexes are constrained by their minimal achievable error bounds, which limits their precision under fixed memory budgets.
}

\vspace{1mm}

\begin{tcolorbox}
\noindent {\bf Observation 2}: The improvement in read performance becomes increasingly marginal as the memory budget grows, suggesting that allocating additional memory to the index may not always yield proportional benefits to the system.
\end{tcolorbox}

\noindent{While increasing the memory budget for indexes generally enhances performance, we observe diminishing returns as index size continues to grow. As shown in Figure~\ref{fig:pointquery}(A), lookup latency decreases substantially when the position boundary is reduced from 256 to 128 and then to 64. However, beyond this point, further reductions in the position boundary yield little to no performance improvement for most learned indexes, even though their memory usage grows exponentially, as shown in Figure~\ref{fig:pointquery}(B).
This phenomenon occurs because, initially, additional memory allows the index to model the key space with higher precision, effectively narrowing the search range and reducing I/O cost. But once the precision reaches the granularity of one or two I/O blocks, the dominant cost—disk I/O—can no longer be significantly reduced. At this stage, further memory investment offers minimal performance gain.

As a result, the growth in memory consumption does not translate into a proportional improvement in lookup latency. This trend mirrors findings from prior benchmarks on learned indexes in in-memory systems~\cite{sosd-vldb}, underscoring a fundamental characteristic of the memory-latency tradeoff inherent to learned index designs.}

\subsection{Impact of Index Granularity}
\label{subsec:indexgranularity}


\begin{tcolorbox}
\noindent {\bf Observation 3}: Learned indexes consume less memory as the granularity grows while it does not significantly affect the query performance.
\end{tcolorbox}

\begin{figure*}[t]
    \setlength{\belowcaptionskip}{-0.4cm}
    \begin{minipage}[b]{.5\textwidth}
    \hspace{-2.5mm}
    \includegraphics[width=\linewidth]{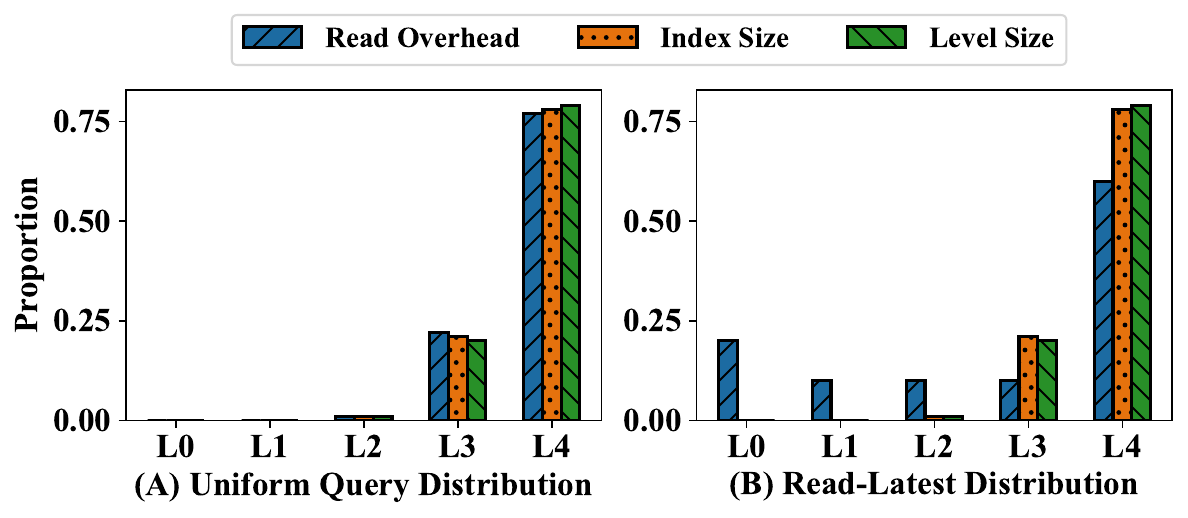}
    \end{minipage}%
\begin{minipage}[b]
{.5\textwidth}
\vspace{2.5mm}
\captionsetup{singlelinecheck = false, justification=raggedright, margin={2mm, 1mm}}
\captionof{table}{The table presents the processing time for each stage of point lookup in PLR, with the position boundary set to 10.
}
\vspace{-2.5mm}
\label{tab:QueryProcess}
\vspace{-1mm}
\raisebox{18mm}{
\hspace{1.5mm}
\renewcommand{\arraystretch}{1}
\scalebox{1.05}{
    \begin{tabular}{|c|c|c|c|}
    \hline
        \multicolumn{4}{|c|}{Statistics of Point Lookup In Detail}\\
        \hline
        Process & SST=4MB & SST=32MB & SST=128MB \\
        \hline
        Table Lookup & 0.19 us / op & 0.11 op / us & 0.07 us / op\\
        Prediction &  0.17 us / op & 0.15 us / op & 0.15 us / op\\ 
        Disk I/O & 2.12 us / op & 2.10 us / op & 2.16 us / op\\
        Binary Search & 0.16 us / op & 0.15 us / op & 0.16 us / op\\
        \hline
    \end{tabular}
}
}
\end{minipage}
\caption{The figure (left) shows the read overhead, the index size, and entries at different levels; the table (right) presents the detail of point lookup.}
\label{fig:readsize}
\end{figure*}
\begin{figure*}
    \centering
    \includegraphics[width=\linewidth]{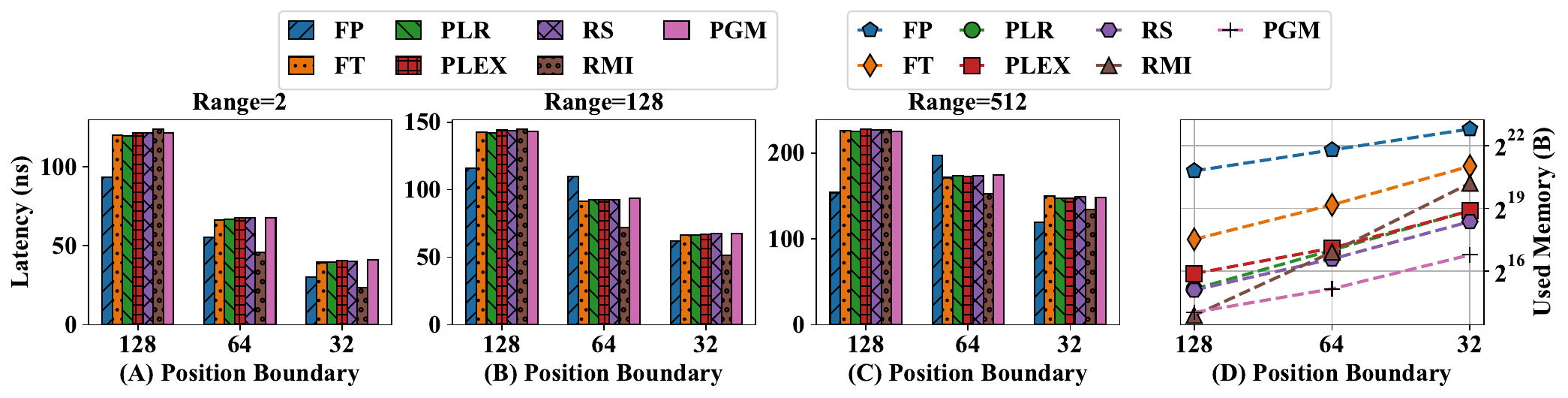}
    \caption{Performance of range lookup under different lookup range and position boundary.}
    \label{fig:range}
\end{figure*}

\noindent{To evaluate the impact of index granularity, we vary the SSTable size from 8MB to 128MB and also include the level-granularity model proposed by Dai \etal~\cite{dai2020wisckey}. We run a point lookup-only workload with 1,000,000 queries and record the latency and memory usage of each index configuration. As shown in Figure~\ref{fig:granularity} (rightmost), lookup latency remains largely unaffected by granularity, varying by only a few microseconds across all configurations. In contrast, granularity has a significant impact on memory usage: increasing SSTable size (i.e., coarser granularity) leads to substantial memory savings, with over a 10× reduction in memory usage when moving from 8MB SSTables to the level model.

Specifically, when the position boundary is greater than 64, memory usage decreases exponentially as SSTable size increases. This is because larger SSTables reduce the total number of SSTables in the system (for a fixed dataset size), thereby lowering the number of inner indexes and associated overhead. However, when the position boundary is smaller than 32, memory usage becomes relatively stable for most learned indexes—except for RMI. In this regime, although fewer inner indexes are needed, the number of segments created by the models remains high, and segment-level overhead becomes the dominant contributor to memory consumption.
RMI behaves differently: its memory usage consistently decreases regardless of the position boundary. This is because RMI's memory cost is primarily driven by its inner index (i.e., the first-stage model), which remains the dominant component of memory usage across configurations, regardless of how many segments are created.

}

\subsection{Compaction Overhead of Learned Indexes}
\begin{tcolorbox}
\noindent {\bf Observation 4}: Unexpectedly, the learning overhead introduced by learned indexes can be considered modest compared to the overall compaction overhead. Index learning and writing time account for less than 5\% of the total compaction time in most of the cases.
\end{tcolorbox}

\vspace{-2mm}
\noindent As discussed in Section \ref{sec:background}, compaction is triggered when a level reaches its maximum capacity, merging the data from that level into the next. In some systems\cite{rocksdb,leveldb}, only a portion of the data is merged to the next level to reduce compaction overhead and prevent write stalls. The compaction process involves reading the sorted run into memory, merging the data, writing it back to disk, and building an index for the new sorted run. When using a learned index, as described in the implementation section, this process also includes segmenting the sorted run and training a model for each segment. Previous research~\cite{dai2020wisckey} highlights that the training time for learned indexes introduces significant overhead, particularly in write-heavy workloads. Therefore, in this section, we evaluate the overhead of learned indexes in a write-only workload with 1,000,000 operations with setting the write buffer to 64MB.

As shown in Figure ~\ref{fig:compaction}, the compaction time remains almost unchanged as the index size grows for all the indexes. This is because the primary cost of compaction comes from reading and writing key-value pairs to and from disk. Additionally, the compaction time for learned indexes does not significantly exceed that of fence pointers. Specifically, most learned indexes show less than a 5\% increase in compaction time, while PLEX exhibits around a 10\% increase.

To break this down, we examine the time spent on learning and writing the model. For most learned indexes, model training takes less than 5\% of the compaction time, which explains the minor increase in total compaction time. However, PLEX uses around 10-15\% of the compaction time for training due to its self-tuning algorithm. As for model writing, most learned indexes consume less than 5\% of compaction time. This demonstrates that the overhead introduced by learned indexes during compaction is modest.

The lower training overhead compared to Bourbon~\cite{dai2020wisckey} result from advancements in hardware. Since the training process occurs in memory, CPU performance is a key factor. Using a high-end CPU, such as the i9-13900K, helps reduce this overhead. Additionally, Bourbon does not pipeline the training process with compaction. Instead, it reads data from the disk first and then performs the training using a background thread, likely leading to higher resource consumption.

\subsection{Read Overhead at Different Levels}
\begin{tcolorbox}
\noindent {\bf Observation 5}: Evenly assigned position boundary across different LSM-tree levels may lead to suboptimal memory-latency tradeoff especially for skewed workload.
\end{tcolorbox}

\vspace{-2mm}
\begin{figure*}
    \centering
    \includegraphics[width=\textwidth]{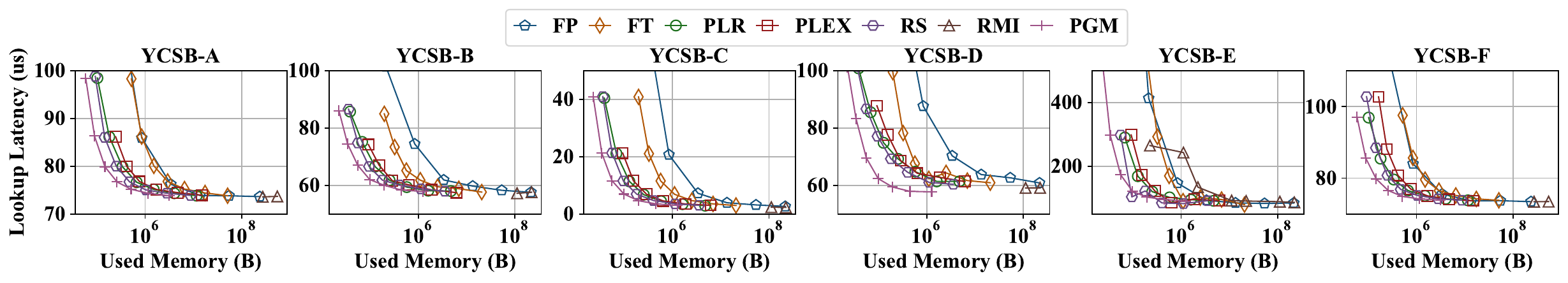}
    \caption{Average operation time of indexes under six YCSB workloads.}
    \label{fig:ycsb}
\end{figure*}
\noindent The LSM-tree has a layered structure, with each level's capacity growing exponentially. This leads to varying read overhead at each level, raising the question: is setting the same position boundary for learned indexes across all levels optimal?

As shown in Figure~\ref{fig:readsize}, we evaluate both uniformly and non-uniformly distributed workloads to measure the read overhead (i.e., lookup time) at each level when using the same position boundary. In the case of uniformly distributed workloads, the proportion of read overhead closely follows the level size, and the index size scales similarly with the level capacity due to the consistent position boundary across levels. However, for skewed workloads (i.e., non-uniform distribution), the read overhead is no longer directly proportional to the level capacity, resulting in an imbalance between index memory allocation and read overhead. 

This imbalance suggests that allocating more memory to levels where read overhead exceeds the index memory proportion could improve the memory-latency tradeoff. In this sense, setting a uniform position boundary across different levels is not always a good choice. This insight is aligned with setting a different memory budget (i.e., bit-per-key) for bloom filters across levels in the LSM-tree~\cite{dayan2017monkey}.

\subsection{Impact on Range Lookup}
\begin{tcolorbox}
\noindent {\bf Observation 6}: Learned indexes offer a superior memory-latency tradeoff compared to fence pointers during short-range lookups. However, this advantage diminishes as the length of the lookup range increases.
\end{tcolorbox}

\noindent Range lookups are a fundamental operation in LSM-tree systems, responsible for retrieving a specified span of key-value pairs. The process begins by locating the starting key of the range at each level, a task facilitated by learned indexes. Once the starting point is identified, the system sequentially retrieves the remaining entries within the target range.

To evaluate the effectiveness of learned indexes for range queries, we test various index types using 64MB SSTables under different range lengths. As shown in Figure~\ref{fig:range}, the memory-latency tradeoff for small ranges resembles that of point lookups. In these cases, all learned indexes outperform the traditional fence pointer in terms of memory efficiency while maintaining comparable query latency.
However, as the range length increases, the latency across different indexing methods begins to converge, reducing the performance advantage of learned indexes. For example, in Figure~\ref{fig:range}(A), increasing the position boundary significantly improves performance in short-range queries. In contrast, as shown in Figure~\ref{fig:range}(C), the same adjustment yields little benefit for longer ranges. This is because, for short ranges, the dominant overhead lies in seeking the initial block, where learned indexes are most effective. In long-range queries, however, the main cost shifts to scanning and retrieving a large number of entries, which diminishes the relative impact of index precision.
As a result, the memory-latency benefits of learned indexes diminish as range length increases, highlighting a limitation of their effectiveness in long-range query scenarios.

\subsection{Impact on Mixed Workload}
\begin{tcolorbox}
\noindent {\bf Observation 7}: The memory-latency tradeoff does not vary a lot under update-lookup mixed workload compared to read-only workloads. Prioritizing read performance is a practical approach when evaluating a learned index in LSM-tree systems.
\end{tcolorbox}

\noindent To evaluate the performance of learned indexes under more complex and real-world conditions, we test them with six YCSB workloads: A is read-write balanced, B is point lookup heavy, C is point lookup only, D focuses on recent point lookup, E is range lookup heavy (with ranges less than 100), and F is read-modify-write (50\% lookup and 50\% update). The result is shown in Figure ~\ref{fig:ycsb}.

Overall, the memory-latency tradeoff remains consistent with the results from the point lookup workloads tested earlier. Specifically, across all workloads, PGM continues to offer the best tradeoff, while FITing-tree lags behind other learned indexes in performance.
\vspace{-3mm}
\section{Discussion}
\label{sec:discussion}
In this section, we are going to conclude about the performance tradeoff of learned indexes on LSM-tree systems and how we can make them suitable for LSM-trees. The evaluation and analysis in the previous section have already found several observations regarding learned indexes in various aspects in the LSM-tree systems. To conclude, learned indexes showcase satisfactory memory-latency tradeoff in both range lookup and point lookup--using less memory and delivering stronger throughput, while the extra computation consumption for training is not remarkable, which demonstrates a superior potential of learned indexes to integrate learned indexes into LSM-tree systems. To help users understand more about having the best learned indexes in their LSM-tree systems, we are going to give three decision guidelines in the following.

\subsection{Insights and Tuning Guide}
\noindent{\bf Prioritize Position Boundary.} In Section ~\ref{sec:config}, we identify three key factors that influence the performance of learned indexes in LSM-tree systems, with position boundary having the most significant impact on read performance while different index types only impact the memory-latency tradeoff. Though some learned indexes focus on optimizing prediction or segment lookup, which can show excellent results in memory-based systems, these advantages become less noticeable in LSM-tree environments. For example, even though PLR employs the simplest inner index structure, its memory-latency tradeoff remains desirable in most cases. This underscores that position boundary plays a more critical role in overall performance compared to prediction optimizations, suggesting that optimizing the model position boundary to a smaller range within the same memory budget is more important than improving the efficiency of indexing these learned models in LSM-trees.

\noindent{\bf Increase the Index Granularity.} While index granularity (i.e., SSTable size) has less impact on performance than the position boundary, increasing granularity can still yield up to a 10\% improvement in the memory-performance tradeoff across all learned indexes. By reducing the memory required to store index structures, larger SSTables make it possible to allocate more memory toward decreasing the position boundary, thereby enhancing lookup performance.
However, adopting a level-granularity model must be approached with caution, as it is only feasible when full merges are performed (i.e., merging an entire level into the next). Although full merges do not increase the overall write amplification~\cite{dayanspooky}, they can lead to short-term spikes in resource usage, which may temporarily degrade foreground performance.

\noindent{\bf Wisely Allocate the Memory Budget.} Since the performance gains diminish once the index size surpasses a certain threshold, allocating excessive memory to learned indexes is not optimal. Instead, allocate a balanced portion of memory to the indexes based on your total budget, and dedicate the remainder to other in-memory components, such as Bloom filters and write buffers, to enhance overall performance. Additionally, because read overhead at each level is not always proportional to the level's capacity, using a uniform position boundary across all levels may not be the best approach. It is more effective to adjust the boundaries according to the specific query distribution.
\vspace{-0mm}
\subsection{Future Direction}
Simply integrating existing learned indexes into LSM-tree systems is not the final step in learned index research. Based on our findings, there are two promising directions for further exploration: (1) developing a more sophisticated algorithm for dynamically allocating memory budgets for learned indexes, taking into account workloads, query distribution, and dataset characteristics to optimize the memory-latency tradeoff, and (2) incorporating learned indexes into the broader optimization of LSM-tree design space, such as in systems like Dostoevsky~\cite{dostoevsky2018}, Wacky~\cite{lsmbush}, and Moose~\cite{moose}. The first direction arises from our observation of the imbalance between read overhead and memory consumption at different levels. The second direction addresses a gap in recent LSM-tree design studies, which largely overlook the role of indexes, especially learned indexes. Given the strong memory-latency tradeoff demonstrated by learned indexes, considering them in the LSM-tree design configuration may lead to valuable optimization insights.

\vspace{0mm}
\section{Related Work}
\label{sec:related}
\noindent{\bf LSM-tree Stores.} Extensive research has focused on optimizing LSM-tree stores through comprehensive theoretical analysis and parameter tuning, such as size ratio, compaction policies, and Bloom filters~\cite{dayan2017monkey,dostoevsky2018,dayanspooky,lsmbush,huynh2021endure,huynh2024towards,moose,ruskey,luo2020rosetta}. These studies have significantly improved the performance of LSM-tree systems. Additionally, works like Dostoevsky~\cite{dostoevsky2018}, Wacky~\cite{lsmbush}, and Moose~\cite{moose} define distinctive LSM-tree structures and derive optimal configurations by theoretically modeling the cost of various LSM-tree operations. Integrating learned indexes into these designs could offer valuable insights. Furthermore, self-tuning systems such as Cosine~\cite{chatterjee2021cosine}, Data Calculator~\cite{datacalculator}, Design Continuums~\cite{idreos2019design}, and Limousine~\cite{chatterjee2024limousine} model the costs of different index structures, including learned indexes and LSM-tree indexes, to calculate the optimal storage structure within a given budget. While these works provide broad insights into storage design, they lack fine-grained guidelines specifically tailored for LSM-tree storage and learned indexes, with some focusing primarily on cloud storage~\cite{chatterjee2024limousine,chatterjee2021cosine}. Our study aims to complement this research by offering more targeted design insights for LSM-tree systems and learned indexes.

\noindent{\bf Learned Indexes.} Our study lies in the improvement of learned indexes techniques. In addition to the learned indexes discussed earlier, we review other notable approaches. MADEX~\cite{hadian2020madex} redesigns B+-tree nodes, incorporating CDF and correction models to enhance point lookups. RUSLI~\cite{hadian2020madex} modifies RadixSpline~\cite{kipf2020radixspline} to support updates, while FINEdex~\cite{li2021finedex} introduces a buffer, building on XIndex~\cite{tang2020xindex}, to handle updates more efficiently. LSI~\cite{kipf2022lsi} is the first to model unsorted data. Some learned indexes, such as AULID~\cite{lan2023simple}, are specifically designed for disk-based systems. Additionally, recent studies have applied learned indexing to string, spatial, and multi-dimensional queries~\cite{ding2020tsunami,li2020lisa,multilearned,qi2020effectively,spector2021bounding,wang2020sindex,gu2023rlr,wu2022nfl}. Several evaluations~\cite{updatablelearnedindex,learnedindexbenchmark,updatableindexready} and surveys~\cite{liu2022survey,ge2023cutting} provide valuable insights into tuning issues and the evolving landscape of learned indexes. Integrating a wider variety of learned indexes into LSM-trees in the future could provide us with more valuable insights.

\noindent{\bf Learned Indexes in LSM-tree Systems.} Due to the compatibility between learned indexes and LSM-tree systems, and the promising memory-latency tradeoff they offer, several recent studies~\cite{googlelearned, dai2020wisckey, empricalleveldb, sarkar2022dissecting, sarkar2023lsm, lu2021tridentkv} explore integrating learned indexes into LSM-trees to improve lookup performance. Abu-Libdeh \etal~\cite{googlelearned} are the first to evaluate the feasibility of learned indexes in LSM-tree systems, though their study does not fully cover different configuration options or index types. Dai \etal~\cite{dai2020wisckey} integrate piecewise linear regression models into their LSM-tree system~\cite{lu2017wisckey} and propose Bourbon, achieving significant lookup improvements. Lu \etal~\cite{lu2021tridentkv} propose TridentKV, which integrates RMI~\cite{rmi} as the learned index and claims better performance than Bourbon in read-heavy workloads. Ramadhan \etal~\cite{empricalleveldb} further improve Bourbon by replacing binary search with exponential search, yielding moderate performance gains. However, these works do not fully explore the entire configuration space that affects learned index performance in LSM-trees, nor do they thoroughly investigate the memory-latency tradeoff. Our work aims to bridge this gap, providing additional insights and extending these foundational studies.

\section{Conclusion}
\label{sec:conclusion}
In this study, we have conducted a comprehensive theoretical and practical evaluation of integrating learned indexes into LSM-tree systems. We begin by revisiting existing learned indexes and analyzing their expected costs to identify key factors that influence LSM-tree performance. Through rigorous evaluations under various conditions, we have derived several design guidelines tailored to LSM-tree systems and provide practical insights for optimizing their performance.

\section{Artifacts}
To facilitate reproducibility and further exploration, we provide the full implementation, including source code, workload generators, and experiment scripts, in our public GitHub repository:
\url{https://github.com/qingshanlanshan/LearnedIndexInLSM}.
The repository includes detailed instructions on how to configure, build, and run the experiments described in this paper. Please refer to the \texttt{README.md} file in the repository for setup instructions, system requirements, and usage examples.

\bibliographystyle{ACM-Reference-Format}
\bibliography{sample-base}


\begin{thebibliography}{56}


\ifx \showCODEN    \undefined \def \showCODEN     #1{\unskip}     \fi
\ifx \showDOI      \undefined \def \showDOI       #1{#1}\fi
\ifx \showISBNx    \undefined \def \showISBNx     #1{\unskip}     \fi
\ifx \showISBNxiii \undefined \def \showISBNxiii  #1{\unskip}     \fi
\ifx \showISSN     \undefined \def \showISSN      #1{\unskip}     \fi
\ifx \showLCCN     \undefined \def \showLCCN      #1{\unskip}     \fi
\ifx \shownote     \undefined \def \shownote      #1{#1}          \fi
\ifx \showarticletitle \undefined \def \showarticletitle #1{#1}   \fi
\ifx \showURL      \undefined \def \showURL       {\relax}        \fi
\providecommand\bibfield[2]{#2}
\providecommand\bibinfo[2]{#2}
\providecommand\natexlab[1]{#1}
\providecommand\showeprint[2][]{arXiv:#2}

\bibitem[tec( )]%
        {technicalreport}
 \bibinfo{year}{-}\natexlab{}.
\newblock \bibinfo{title}{Learned-Index-for-LSM-tree technical report}.
\newblock \bibinfo{howpublished}{\url{https://github.com/qingshanlanshan/LearnedIndexInLSM/TechnicalReport.pdf}}.
\newblock


\bibitem[Abu-Libdeh et~al\mbox{.}(2020)]%
        {googlelearned}
\bibfield{author}{\bibinfo{person}{Hussam Abu-Libdeh}, \bibinfo{person}{Deniz Alt{\i}nb{\"u}ken}, \bibinfo{person}{Alex Beutel}, \bibinfo{person}{Ed~H Chi}, \bibinfo{person}{Lyric Doshi}, \bibinfo{person}{Tim Kraska}, \bibinfo{person}{Andy Ly}, \bibinfo{person}{Christopher Olston}, {et~al\mbox{.}}} \bibinfo{year}{2020}\natexlab{}.
\newblock \showarticletitle{Learned indexes for a google-scale disk-based database}.
\newblock \bibinfo{journal}{\emph{arXiv preprint arXiv:2012.12501}} (\bibinfo{year}{2020}).
\newblock


\bibitem[Chatterjee et~al\mbox{.}(2021)]%
        {chatterjee2021cosine}
\bibfield{author}{\bibinfo{person}{Subarna Chatterjee}, \bibinfo{person}{Meena Jagadeesan}, \bibinfo{person}{Wilson Qin}, {and} \bibinfo{person}{Stratos Idreos}.} \bibinfo{year}{2021}\natexlab{}.
\newblock \showarticletitle{Cosine: a cloud-cost optimized self-designing key-value storage engine}.
\newblock \bibinfo{journal}{\emph{Proceedings of the VLDB Endowment}} \bibinfo{volume}{15}, \bibinfo{number}{1} (\bibinfo{year}{2021}), \bibinfo{pages}{112--126}.
\newblock


\bibitem[Chatterjee et~al\mbox{.}(2024)]%
        {chatterjee2024limousine}
\bibfield{author}{\bibinfo{person}{Subarna Chatterjee}, \bibinfo{person}{Mark~F Pekala}, \bibinfo{person}{Lev Kruglyak}, {and} \bibinfo{person}{Stratos Idreos}.} \bibinfo{year}{2024}\natexlab{}.
\newblock \showarticletitle{Limousine: Blending Learned and Classical Indexes to Self-Design Larger-than-Memory Cloud Storage Engines}.
\newblock \bibinfo{journal}{\emph{Proceedings of the ACM on Management of Data}} \bibinfo{volume}{2}, \bibinfo{number}{1} (\bibinfo{year}{2024}), \bibinfo{pages}{1--28}.
\newblock


\bibitem[Code(2024)]%
        {wiredtiger}
\bibfield{author}{\bibinfo{person}{Source Code}.} \bibinfo{year}{2024}\natexlab{}.
\newblock \bibinfo{title}{WiredTiger}.
\newblock \bibinfo{howpublished}{\url{https://github.com/wiredtiger/wiredtiger}}.
\newblock


\bibitem[Corbett et~al\mbox{.}(2013)]%
        {corbett2013spanner}
\bibfield{author}{\bibinfo{person}{James~C Corbett}, \bibinfo{person}{Jeffrey Dean}, \bibinfo{person}{Michael Epstein}, \bibinfo{person}{Andrew Fikes}, \bibinfo{person}{Christopher Frost}, \bibinfo{person}{Jeffrey~John Furman}, \bibinfo{person}{Sanjay Ghemawat}, \bibinfo{person}{Andrey Gubarev}, \bibinfo{person}{Christopher Heiser}, \bibinfo{person}{Peter Hochschild}, {et~al\mbox{.}}} \bibinfo{year}{2013}\natexlab{}.
\newblock \showarticletitle{Spanner: Google’s globally distributed database}.
\newblock \bibinfo{journal}{\emph{ACM Transactions on Computer Systems (TOCS)}} \bibinfo{volume}{31}, \bibinfo{number}{3} (\bibinfo{year}{2013}), \bibinfo{pages}{1--22}.
\newblock


\bibitem[Dai et~al\mbox{.}(2020)]%
        {dai2020wisckey}
\bibfield{author}{\bibinfo{person}{Yifan Dai}, \bibinfo{person}{Yien Xu}, \bibinfo{person}{Aishwarya Ganesan}, \bibinfo{person}{Ramnatthan Alagappan}, \bibinfo{person}{Brian Kroth}, \bibinfo{person}{Andrea Arpaci-Dusseau}, {and} \bibinfo{person}{Remzi Arpaci-Dusseau}.} \bibinfo{year}{2020}\natexlab{}.
\newblock \showarticletitle{From WiscKey to Bourbon: A Learned Index for Log-Structured Merge Trees}. In \bibinfo{booktitle}{\emph{14th USENIX Symposium on Operating Systems Design and Implementation (OSDI 20)}}. \bibinfo{pages}{155--171}.
\newblock


\bibitem[Dayan et~al\mbox{.}(2017)]%
        {dayan2017monkey}
\bibfield{author}{\bibinfo{person}{Niv Dayan}, \bibinfo{person}{Manos Athanassoulis}, {and} \bibinfo{person}{Stratos Idreos}.} \bibinfo{year}{2017}\natexlab{}.
\newblock \showarticletitle{Monkey: Optimal navigable key-value store}. In \bibinfo{booktitle}{\emph{Proceedings of the 2017 ACM International Conference on Management of Data}}. \bibinfo{pages}{79--94}.
\newblock


\bibitem[Dayan and Idreos(2018)]%
        {dostoevsky2018}
\bibfield{author}{\bibinfo{person}{Niv Dayan} {and} \bibinfo{person}{Stratos Idreos}.} \bibinfo{year}{2018}\natexlab{}.
\newblock \showarticletitle{Dostoevsky: Better Space-Time Trade-Offs for LSM-Tree Based Key-Value Stores via Adaptive Removal of Superfluous Merging}. In \bibinfo{booktitle}{\emph{Proceedings of the 2018 International Conference on Management of Data}} (Houston, TX, USA) \emph{(\bibinfo{series}{SIGMOD '18})}. \bibinfo{publisher}{Association for Computing Machinery}, \bibinfo{address}{New York, NY, USA}, \bibinfo{pages}{505–520}.
\newblock
\showISBNx{9781450347037}
\urldef\tempurl%
\url{https://doi.org/10.1145/3183713.3196927}
\showDOI{\tempurl}


\bibitem[Dayan and Idreos(2019)]%
        {lsmbush}
\bibfield{author}{\bibinfo{person}{Niv Dayan} {and} \bibinfo{person}{Stratos Idreos}.} \bibinfo{year}{2019}\natexlab{}.
\newblock \showarticletitle{The log-structured merge-bush \& the wacky continuum}. In \bibinfo{booktitle}{\emph{Proceedings of the 2019 International Conference on Management of Data}}. \bibinfo{pages}{449--466}.
\newblock


\bibitem[Dayan et~al\mbox{.}(2022)]%
        {dayanspooky}
\bibfield{author}{\bibinfo{person}{Niv Dayan}, \bibinfo{person}{Tamar Weiss}, \bibinfo{person}{Shmuel Dashevsky}, \bibinfo{person}{Michael Pan}, \bibinfo{person}{Edward Bortnikov}, {and} \bibinfo{person}{Moshe Twitto}.} \bibinfo{year}{2022}\natexlab{}.
\newblock \showarticletitle{Spooky: granulating LSM-tree compactions correctly}.
\newblock \bibinfo{journal}{\emph{Proceedings of the VLDB Endowment}} \bibinfo{volume}{15}, \bibinfo{number}{11} (\bibinfo{year}{2022}), \bibinfo{pages}{3071--3084}.
\newblock


\bibitem[Ding et~al\mbox{.}(2020a)]%
        {ding2020alex}
\bibfield{author}{\bibinfo{person}{Jialin Ding}, \bibinfo{person}{Umar~Farooq Minhas}, \bibinfo{person}{Jia Yu}, \bibinfo{person}{Chi Wang}, \bibinfo{person}{Jaeyoung Do}, \bibinfo{person}{Yinan Li}, \bibinfo{person}{Hantian Zhang}, \bibinfo{person}{Badrish Chandramouli}, \bibinfo{person}{Johannes Gehrke}, \bibinfo{person}{Donald Kossmann}, {et~al\mbox{.}}} \bibinfo{year}{2020}\natexlab{a}.
\newblock \showarticletitle{ALEX: an updatable adaptive learned index}. In \bibinfo{booktitle}{\emph{Proceedings of the 2020 ACM SIGMOD International Conference on Management of Data}}. \bibinfo{pages}{969--984}.
\newblock


\bibitem[Ding et~al\mbox{.}(2020b)]%
        {ding2020tsunami}
\bibfield{author}{\bibinfo{person}{Jialin Ding}, \bibinfo{person}{Vikram Nathan}, \bibinfo{person}{Mohammad Alizadeh}, {and} \bibinfo{person}{Tim Kraska}.} \bibinfo{year}{2020}\natexlab{b}.
\newblock \showarticletitle{Tsunami: A learned multi-dimensional index for correlated data and skewed workloads}.
\newblock \bibinfo{journal}{\emph{arXiv preprint arXiv:2006.13282}} (\bibinfo{year}{2020}).
\newblock


\bibitem[Facebook(2024)]%
        {rocksdb}
\bibfield{author}{\bibinfo{person}{Facebook}.} \bibinfo{year}{2024}\natexlab{}.
\newblock \bibinfo{title}{RocksDB}.
\newblock \bibinfo{howpublished}{\url{https://github.com/facebook/rocksdb}}.
\newblock


\bibitem[Ferragina and Vinciguerra(2020)]%
        {ferragina2020pgm}
\bibfield{author}{\bibinfo{person}{Paolo Ferragina} {and} \bibinfo{person}{Giorgio Vinciguerra}.} \bibinfo{year}{2020}\natexlab{}.
\newblock \showarticletitle{The PGM-index: a fully-dynamic compressed learned index with provable worst-case bounds}.
\newblock \bibinfo{journal}{\emph{Proceedings of the VLDB Endowment}} \bibinfo{volume}{13}, \bibinfo{number}{8} (\bibinfo{year}{2020}), \bibinfo{pages}{1162--1175}.
\newblock


\bibitem[Galakatos et~al\mbox{.}(2019)]%
        {fitingtree}
\bibfield{author}{\bibinfo{person}{Alex Galakatos}, \bibinfo{person}{Michael Markovitch}, \bibinfo{person}{Carsten Binnig}, \bibinfo{person}{Rodrigo Fonseca}, {and} \bibinfo{person}{Tim Kraska}.} \bibinfo{year}{2019}\natexlab{}.
\newblock \showarticletitle{Fiting-tree: A data-aware index structure}. In \bibinfo{booktitle}{\emph{Proceedings of the 2019 international conference on management of data}}. \bibinfo{pages}{1189--1206}.
\newblock


\bibitem[Ge et~al\mbox{.}(2023)]%
        {ge2023cutting}
\bibfield{author}{\bibinfo{person}{Jiake Ge}, \bibinfo{person}{Boyu Shi}, \bibinfo{person}{Yanfeng Chai}, \bibinfo{person}{Yuanhui Luo}, \bibinfo{person}{Yunda Guo}, \bibinfo{person}{Yinxuan He}, {and} \bibinfo{person}{Yunpeng Chai}.} \bibinfo{year}{2023}\natexlab{}.
\newblock \showarticletitle{Cutting Learned Index into Pieces: An In-depth Inquiry into Updatable Learned Indexes}. In \bibinfo{booktitle}{\emph{2023 IEEE 39th International Conference on Data Engineering (ICDE)}}. IEEE, \bibinfo{pages}{315--327}.
\newblock


\bibitem[Google(2024)]%
        {leveldb}
\bibfield{author}{\bibinfo{person}{Google}.} \bibinfo{year}{2024}\natexlab{}.
\newblock \bibinfo{title}{LevelDB}.
\newblock \bibinfo{howpublished}{\url{https://github.com/google/leveldb/}}.
\newblock


\bibitem[Gu et~al\mbox{.}(2023)]%
        {gu2023rlr}
\bibfield{author}{\bibinfo{person}{Tu Gu}, \bibinfo{person}{Kaiyu Feng}, \bibinfo{person}{Gao Cong}, \bibinfo{person}{Cheng Long}, \bibinfo{person}{Zheng Wang}, {and} \bibinfo{person}{Sheng Wang}.} \bibinfo{year}{2023}\natexlab{}.
\newblock \showarticletitle{The rlr-tree: A reinforcement learning based r-tree for spatial data}.
\newblock \bibinfo{journal}{\emph{Proceedings of the ACM on Management of Data}} \bibinfo{volume}{1}, \bibinfo{number}{1} (\bibinfo{year}{2023}), \bibinfo{pages}{1--26}.
\newblock


\bibitem[Hadian and Heinis(2020)]%
        {hadian2020madex}
\bibfield{author}{\bibinfo{person}{Ali Hadian} {and} \bibinfo{person}{Thomas Heinis}.} \bibinfo{year}{2020}\natexlab{}.
\newblock \showarticletitle{MADEX: Learning-augmented Algorithmic Index Structures.}. In \bibinfo{booktitle}{\emph{AIDB@ VLDB}}.
\newblock


\bibitem[Huynh et~al\mbox{.}(2021)]%
        {huynh2021endure}
\bibfield{author}{\bibinfo{person}{Andy Huynh}, \bibinfo{person}{Harshal Chaudhari}, \bibinfo{person}{Evimaria Terzi}, {and} \bibinfo{person}{Manos Athanassoulis}.} \bibinfo{year}{2021}\natexlab{}.
\newblock \showarticletitle{Endure: A Robust Tuning Paradigm for LSM Trees Under Workload Uncertainty}.
\newblock \bibinfo{journal}{\emph{arXiv preprint arXiv:2110.13801}} (\bibinfo{year}{2021}).
\newblock


\bibitem[Huynh et~al\mbox{.}(2024)]%
        {huynh2024towards}
\bibfield{author}{\bibinfo{person}{Andy Huynh}, \bibinfo{person}{Harshal~A Chaudhari}, \bibinfo{person}{Evimaria Terzi}, {and} \bibinfo{person}{Manos Athanassoulis}.} \bibinfo{year}{2024}\natexlab{}.
\newblock \showarticletitle{Towards flexibility and robustness of LSM trees}.
\newblock \bibinfo{journal}{\emph{The VLDB Journal}} (\bibinfo{year}{2024}), \bibinfo{pages}{1--24}.
\newblock


\bibitem[Idreos et~al\mbox{.}(2019)]%
        {idreos2019design}
\bibfield{author}{\bibinfo{person}{Stratos Idreos}, \bibinfo{person}{Niv Dayan}, \bibinfo{person}{Wilson Qin}, \bibinfo{person}{Mali Akmanalp}, \bibinfo{person}{Sophie Hilgard}, \bibinfo{person}{Andrew Ross}, \bibinfo{person}{James Lennon}, \bibinfo{person}{Varun Jain}, \bibinfo{person}{Harshita Gupta}, \bibinfo{person}{David Li}, {et~al\mbox{.}}} \bibinfo{year}{2019}\natexlab{}.
\newblock \showarticletitle{Design Continuums and the Path Toward Self-Designing Key-Value Stores that Know and Learn.}. In \bibinfo{booktitle}{\emph{CIDR}}.
\newblock


\bibitem[Idreos et~al\mbox{.}(2018)]%
        {datacalculator}
\bibfield{author}{\bibinfo{person}{Stratos Idreos}, \bibinfo{person}{Kostas Zoumpatianos}, \bibinfo{person}{Brian Hentschel}, \bibinfo{person}{Michael~S Kester}, {and} \bibinfo{person}{Demi Guo}.} \bibinfo{year}{2018}\natexlab{}.
\newblock \showarticletitle{The data calculator: Data structure design and cost synthesis from first principles and learned cost models}. In \bibinfo{booktitle}{\emph{Proceedings of the 2018 International Conference on Management of Data}}. \bibinfo{pages}{535--550}.
\newblock


\bibitem[Kipf et~al\mbox{.}(2022)]%
        {kipf2022lsi}
\bibfield{author}{\bibinfo{person}{Andreas Kipf}, \bibinfo{person}{Dominik Horn}, \bibinfo{person}{Pascal Pfeil}, \bibinfo{person}{Ryan Marcus}, {and} \bibinfo{person}{Tim Kraska}.} \bibinfo{year}{2022}\natexlab{}.
\newblock \showarticletitle{LSI: a learned secondary index structure}. In \bibinfo{booktitle}{\emph{Proceedings of the Fifth International Workshop on Exploiting Artificial Intelligence Techniques for Data Management}}. \bibinfo{pages}{1--5}.
\newblock


\bibitem[Kipf et~al\mbox{.}(2020)]%
        {kipf2020radixspline}
\bibfield{author}{\bibinfo{person}{Andreas Kipf}, \bibinfo{person}{Ryan Marcus}, \bibinfo{person}{Alexander van Renen}, \bibinfo{person}{Mihail Stoian}, \bibinfo{person}{Alfons Kemper}, \bibinfo{person}{Tim Kraska}, {and} \bibinfo{person}{Thomas Neumann}.} \bibinfo{year}{2020}\natexlab{}.
\newblock \showarticletitle{RadixSpline: a single-pass learned index}. In \bibinfo{booktitle}{\emph{Proceedings of the third international workshop on exploiting artificial intelligence techniques for data management}}. \bibinfo{pages}{1--5}.
\newblock


\bibitem[Kraska et~al\mbox{.}(2018a)]%
        {kraska2018learnedIndex}
\bibfield{author}{\bibinfo{person}{Tim Kraska}, \bibinfo{person}{Alex Beutel}, \bibinfo{person}{Ed~H Chi}, \bibinfo{person}{Jeffrey Dean}, {and} \bibinfo{person}{Neoklis Polyzotis}.} \bibinfo{year}{2018}\natexlab{a}.
\newblock \showarticletitle{The case for learned index structures}. In \bibinfo{booktitle}{\emph{Proceedings of the 2018 international conference on management of data}}. \bibinfo{pages}{489--504}.
\newblock


\bibitem[Kraska et~al\mbox{.}(2018b)]%
        {rmi}
\bibfield{author}{\bibinfo{person}{Tim Kraska}, \bibinfo{person}{Alex Beutel}, \bibinfo{person}{Ed~H Chi}, \bibinfo{person}{Jeffrey Dean}, {and} \bibinfo{person}{Neoklis Polyzotis}.} \bibinfo{year}{2018}\natexlab{b}.
\newblock \showarticletitle{The case for learned index structures}. In \bibinfo{booktitle}{\emph{Proceedings of the 2018 international conference on management of data}}. \bibinfo{pages}{489--504}.
\newblock


\bibitem[Lakshman and Malik(2010)]%
        {lakshman2010cassandra}
\bibfield{author}{\bibinfo{person}{Avinash Lakshman} {and} \bibinfo{person}{Prashant Malik}.} \bibinfo{year}{2010}\natexlab{}.
\newblock \showarticletitle{Cassandra: a decentralized structured storage system}.
\newblock \bibinfo{journal}{\emph{ACM SIGOPS Operating Systems Review}} \bibinfo{volume}{44}, \bibinfo{number}{2} (\bibinfo{year}{2010}), \bibinfo{pages}{35--40}.
\newblock


\bibitem[Lan et~al\mbox{.}(2023a)]%
        {updatablelearnedindex}
\bibfield{author}{\bibinfo{person}{Hai Lan}, \bibinfo{person}{Zhifeng Bao}, \bibinfo{person}{J.~Shane Culpepper}, {and} \bibinfo{person}{Renata Borovica-Gajic}.} \bibinfo{year}{2023}\natexlab{a}.
\newblock \showarticletitle{Updatable Learned Indexes Meet Disk-Resident DBMS - From Evaluations to Design Choices}.
\newblock \bibinfo{journal}{\emph{Proc. ACM Manag. Data}} \bibinfo{volume}{1}, \bibinfo{number}{2}, Article \bibinfo{articleno}{139} (\bibinfo{date}{June} \bibinfo{year}{2023}), \bibinfo{numpages}{22}~pages.
\newblock
\urldef\tempurl%
\url{https://doi.org/10.1145/3589284}
\showURL{%
\tempurl}


\bibitem[Lan et~al\mbox{.}(2023b)]%
        {lan2023simple}
\bibfield{author}{\bibinfo{person}{Hai Lan}, \bibinfo{person}{Zhifeng Bao}, \bibinfo{person}{J~Shane Culpepper}, \bibinfo{person}{Renata Borovica-Gajic}, {and} \bibinfo{person}{Yu Dong}.} \bibinfo{year}{2023}\natexlab{b}.
\newblock \showarticletitle{A simple yet high-performing on-disk learned index: Can we have our cake and eat it too?}
\newblock \bibinfo{journal}{\emph{arXiv preprint arXiv:2306.02604}} (\bibinfo{year}{2023}).
\newblock


\bibitem[Li et~al\mbox{.}(2021)]%
        {li2021finedex}
\bibfield{author}{\bibinfo{person}{Pengfei Li}, \bibinfo{person}{Yu Hua}, \bibinfo{person}{Jingnan Jia}, {and} \bibinfo{person}{Pengfei Zuo}.} \bibinfo{year}{2021}\natexlab{}.
\newblock \showarticletitle{FINEdex: a fine-grained learned index scheme for scalable and concurrent memory systems}.
\newblock \bibinfo{journal}{\emph{Proceedings of the VLDB Endowment}} \bibinfo{volume}{15}, \bibinfo{number}{2} (\bibinfo{year}{2021}), \bibinfo{pages}{321--334}.
\newblock


\bibitem[Li et~al\mbox{.}(2020)]%
        {li2020lisa}
\bibfield{author}{\bibinfo{person}{Pengfei Li}, \bibinfo{person}{Hua Lu}, \bibinfo{person}{Qian Zheng}, \bibinfo{person}{Long Yang}, {and} \bibinfo{person}{Gang Pan}.} \bibinfo{year}{2020}\natexlab{}.
\newblock \showarticletitle{LISA: A learned index structure for spatial data}. In \bibinfo{booktitle}{\emph{Proceedings of the 2020 ACM SIGMOD international conference on management of data}}. \bibinfo{pages}{2119--2133}.
\newblock


\bibitem[Li et~al\mbox{.}(2023)]%
        {li2023dili}
\bibfield{author}{\bibinfo{person}{Pengfei Li}, \bibinfo{person}{Hua Lu}, \bibinfo{person}{Rong Zhu}, \bibinfo{person}{Bolin Ding}, \bibinfo{person}{Long Yang}, {and} \bibinfo{person}{Gang Pan}.} \bibinfo{year}{2023}\natexlab{}.
\newblock \showarticletitle{DILI: A Distribution-Driven Learned Index (Extended version)}.
\newblock \bibinfo{journal}{\emph{arXiv preprint arXiv:2304.08817}} (\bibinfo{year}{2023}).
\newblock


\bibitem[Liu et~al\mbox{.}(2024)]%
        {moose}
\bibfield{author}{\bibinfo{person}{Junfeng Liu}, \bibinfo{person}{Fan Wang}, \bibinfo{person}{Dingheng Mo}, {and} \bibinfo{person}{Siqiang Luo}.} \bibinfo{year}{2024}\natexlab{}.
\newblock \showarticletitle{Structural Designs Meet Optimality: Exploring Optimized LSM-tree Structures in A Colossal Configuration Space}.
\newblock \bibinfo{journal}{\emph{Proceedings of the ACM on Management of Data}} \bibinfo{volume}{2}, \bibinfo{number}{3} (\bibinfo{year}{2024}), \bibinfo{pages}{1--26}.
\newblock


\bibitem[Liu et~al\mbox{.}(2022)]%
        {liu2022survey}
\bibfield{author}{\bibinfo{person}{Yu Liu}, \bibinfo{person}{Hua Wang}, \bibinfo{person}{Ke Zhou}, \bibinfo{person}{ChunHua Li}, {and} \bibinfo{person}{Rengeng Wu}.} \bibinfo{year}{2022}\natexlab{}.
\newblock \showarticletitle{A survey on AI for storage}.
\newblock \bibinfo{journal}{\emph{CCF Transactions on High Performance Computing}} \bibinfo{volume}{4}, \bibinfo{number}{3} (\bibinfo{year}{2022}), \bibinfo{pages}{233--264}.
\newblock


\bibitem[Lu et~al\mbox{.}(2021)]%
        {lu2021tridentkv}
\bibfield{author}{\bibinfo{person}{Kai Lu}, \bibinfo{person}{Nannan Zhao}, \bibinfo{person}{Jiguang Wan}, \bibinfo{person}{Changhong Fei}, \bibinfo{person}{Wei Zhao}, {and} \bibinfo{person}{Tongliang Deng}.} \bibinfo{year}{2021}\natexlab{}.
\newblock \showarticletitle{TridentKV: A read-optimized LSM-tree based KV store via adaptive indexing and space-efficient partitioning}.
\newblock \bibinfo{journal}{\emph{IEEE Transactions on Parallel and Distributed Systems}} \bibinfo{volume}{33}, \bibinfo{number}{8} (\bibinfo{year}{2021}), \bibinfo{pages}{1953--1966}.
\newblock


\bibitem[Lu et~al\mbox{.}(2017)]%
        {lu2017wisckey}
\bibfield{author}{\bibinfo{person}{Lanyue Lu}, \bibinfo{person}{Thanumalayan~Sankaranarayana Pillai}, \bibinfo{person}{Hariharan Gopalakrishnan}, \bibinfo{person}{Andrea~C Arpaci-Dusseau}, {and} \bibinfo{person}{Remzi~H Arpaci-Dusseau}.} \bibinfo{year}{2017}\natexlab{}.
\newblock \showarticletitle{Wisckey: Separating keys from values in ssd-conscious storage}.
\newblock \bibinfo{journal}{\emph{ACM Transactions on Storage (TOS)}} \bibinfo{volume}{13}, \bibinfo{number}{1} (\bibinfo{year}{2017}), \bibinfo{pages}{1--28}.
\newblock


\bibitem[Luo et~al\mbox{.}(2020)]%
        {luo2020rosetta}
\bibfield{author}{\bibinfo{person}{Siqiang Luo}, \bibinfo{person}{Subarna Chatterjee}, \bibinfo{person}{Rafael Ketsetsidis}, \bibinfo{person}{Niv Dayan}, \bibinfo{person}{Wilson Qin}, {and} \bibinfo{person}{Stratos Idreos}.} \bibinfo{year}{2020}\natexlab{}.
\newblock \showarticletitle{Rosetta: A robust space-time optimized range filter for key-value stores}. In \bibinfo{booktitle}{\emph{Proceedings of the 2020 ACM SIGMOD International Conference on Management of Data}}. \bibinfo{pages}{2071--2086}.
\newblock


\bibitem[Maltry and Dittrich(2022)]%
        {rmianalytics}
\bibfield{author}{\bibinfo{person}{Marcel Maltry} {and} \bibinfo{person}{Jens Dittrich}.} \bibinfo{year}{2022}\natexlab{}.
\newblock \showarticletitle{A Critical Analysis of Recursive Model Indexes}.
\newblock \bibinfo{journal}{\emph{Proc. {VLDB} Endow.}} \bibinfo{volume}{15}, \bibinfo{number}{5} (\bibinfo{year}{2022}), \bibinfo{pages}{1079--1091}.
\newblock


\bibitem[Marcus et~al\mbox{.}(2020a)]%
        {learnedindexbenchmark}
\bibfield{author}{\bibinfo{person}{Ryan Marcus}, \bibinfo{person}{Andreas Kipf}, \bibinfo{person}{Alexander van Renen}, \bibinfo{person}{Mihail Stoian}, \bibinfo{person}{Sanchit Misra}, \bibinfo{person}{Alfons Kemper}, \bibinfo{person}{Thomas Neumann}, {and} \bibinfo{person}{Tim Kraska}.} \bibinfo{year}{2020}\natexlab{a}.
\newblock \showarticletitle{Benchmarking learned indexes}.
\newblock \bibinfo{journal}{\emph{arXiv preprint arXiv:2006.12804}} (\bibinfo{year}{2020}).
\newblock


\bibitem[Marcus et~al\mbox{.}(2020b)]%
        {sosd-vldb}
\bibfield{author}{\bibinfo{person}{Ryan Marcus}, \bibinfo{person}{Andreas Kipf}, \bibinfo{person}{Alexander van Renen}, \bibinfo{person}{Mihail Stoian}, \bibinfo{person}{Sanchit Misra}, \bibinfo{person}{Alfons Kemper}, \bibinfo{person}{Thomas Neumann}, {and} \bibinfo{person}{Tim Kraska}.} \bibinfo{year}{2020}\natexlab{b}.
\newblock \showarticletitle{Benchmarking Learned Indexes}.
\newblock \bibinfo{journal}{\emph{Proc. {VLDB} Endow.}} \bibinfo{volume}{14}, \bibinfo{number}{1} (\bibinfo{year}{2020}), \bibinfo{pages}{1--13}.
\newblock


\bibitem[Mo et~al\mbox{.}(2023)]%
        {ruskey}
\bibfield{author}{\bibinfo{person}{Dingheng Mo}, \bibinfo{person}{Fanchao Chen}, \bibinfo{person}{Siqiang Luo}, {and} \bibinfo{person}{Caihua Shan}.} \bibinfo{year}{2023}\natexlab{}.
\newblock \showarticletitle{Learning to Optimize LSM-trees: Towards A Reinforcement Learning based Key-Value Store for Dynamic Workloads}.
\newblock \bibinfo{journal}{\emph{Proc. ACM Manag. Data}} \bibinfo{volume}{1}, \bibinfo{number}{3}, Article \bibinfo{articleno}{213} (\bibinfo{date}{Nov.} \bibinfo{year}{2023}), \bibinfo{numpages}{25}~pages.
\newblock
\urldef\tempurl%
\url{https://doi.org/10.1145/3617333}
\showDOI{\tempurl}


\bibitem[Nathan et~al\mbox{.}(2020)]%
        {multilearned}
\bibfield{author}{\bibinfo{person}{Vikram Nathan}, \bibinfo{person}{Jialin Ding}, \bibinfo{person}{Mohammad Alizadeh}, {and} \bibinfo{person}{Tim Kraska}.} \bibinfo{year}{2020}\natexlab{}.
\newblock \showarticletitle{Learning multi-dimensional indexes}. In \bibinfo{booktitle}{\emph{Proceedings of the 2020 ACM SIGMOD international conference on management of data}}. \bibinfo{pages}{985--1000}.
\newblock


\bibitem[Qi et~al\mbox{.}(2020)]%
        {qi2020effectively}
\bibfield{author}{\bibinfo{person}{Jianzhong Qi}, \bibinfo{person}{Guanli Liu}, \bibinfo{person}{Christian~S Jensen}, {and} \bibinfo{person}{Lars Kulik}.} \bibinfo{year}{2020}\natexlab{}.
\newblock \showarticletitle{Effectively learning spatial indices}.
\newblock \bibinfo{journal}{\emph{Proceedings of the VLDB Endowment}} \bibinfo{volume}{13}, \bibinfo{number}{12} (\bibinfo{year}{2020}), \bibinfo{pages}{2341--2354}.
\newblock


\bibitem[Raju et~al\mbox{.}(2017)]%
        {raju2017pebblesdb}
\bibfield{author}{\bibinfo{person}{Pandian Raju}, \bibinfo{person}{Rohan Kadekodi}, \bibinfo{person}{Vijay Chidambaram}, {and} \bibinfo{person}{Ittai Abraham}.} \bibinfo{year}{2017}\natexlab{}.
\newblock \showarticletitle{Pebblesdb: Building key-value stores using fragmented log-structured merge trees}. In \bibinfo{booktitle}{\emph{Proceedings of the 26th Symposium on Operating Systems Principles}}. \bibinfo{pages}{497--514}.
\newblock


\bibitem[Ramadhan et~al\mbox{.}(2023)]%
        {empricalleveldb}
\bibfield{author}{\bibinfo{person}{Agung~Rahmat Ramadhan}, \bibinfo{person}{Min-guk Choi}, \bibinfo{person}{Yoojin Chung}, {and} \bibinfo{person}{Jongmoo Choi}.} \bibinfo{year}{2023}\natexlab{}.
\newblock \showarticletitle{An Empirical Study of Segmented Linear Regression Search in LevelDB}.
\newblock \bibinfo{journal}{\emph{Electronics}} \bibinfo{volume}{12}, \bibinfo{number}{4} (\bibinfo{year}{2023}), \bibinfo{pages}{1018}.
\newblock


\bibitem[Sarkar and Athanassoulis(2022)]%
        {sarkar2022dissecting}
\bibfield{author}{\bibinfo{person}{Subhadeep Sarkar} {and} \bibinfo{person}{Manos Athanassoulis}.} \bibinfo{year}{2022}\natexlab{}.
\newblock \showarticletitle{Dissecting, designing, and optimizing LSM-based data stores}. In \bibinfo{booktitle}{\emph{Proceedings of the 2022 International Conference on Management of Data}}. \bibinfo{pages}{2489--2497}.
\newblock


\bibitem[Sarkar et~al\mbox{.}(2023)]%
        {sarkar2023lsm}
\bibfield{author}{\bibinfo{person}{Subhadeep Sarkar}, \bibinfo{person}{Niv Dayan}, {and} \bibinfo{person}{Manos Athanassoulis}.} \bibinfo{year}{2023}\natexlab{}.
\newblock \showarticletitle{The LSM design space and its read optimizations}. In \bibinfo{booktitle}{\emph{2023 IEEE 39th International Conference on Data Engineering (ICDE)}}. IEEE, \bibinfo{pages}{3578--3584}.
\newblock


\bibitem[Spector et~al\mbox{.}(2021)]%
        {spector2021bounding}
\bibfield{author}{\bibinfo{person}{Benjamin Spector}, \bibinfo{person}{Andreas Kipf}, \bibinfo{person}{Kapil Vaidya}, \bibinfo{person}{Chi Wang}, \bibinfo{person}{Umar~Farooq Minhas}, {and} \bibinfo{person}{Tim Kraska}.} \bibinfo{year}{2021}\natexlab{}.
\newblock \showarticletitle{Bounding the last mile: Efficient learned string indexing}.
\newblock \bibinfo{journal}{\emph{arXiv preprint arXiv:2111.14905}} (\bibinfo{year}{2021}).
\newblock


\bibitem[Stoian et~al\mbox{.}(2021)]%
        {plex}
\bibfield{author}{\bibinfo{person}{Mihail Stoian}, \bibinfo{person}{Andreas Kipf}, \bibinfo{person}{Ryan Marcus}, {and} \bibinfo{person}{Tim Kraska}.} \bibinfo{year}{2021}\natexlab{}.
\newblock \showarticletitle{{PLEX:} Towards Practical Learned Indexing}.
\newblock \bibinfo{journal}{\emph{CoRR}}  \bibinfo{volume}{abs/2108.05117} (\bibinfo{year}{2021}).
\newblock
\showeprint[arXiv]{2108.05117}
\urldef\tempurl%
\url{https://arxiv.org/abs/2108.05117}
\showURL{%
\tempurl}


\bibitem[Tang et~al\mbox{.}(2020)]%
        {tang2020xindex}
\bibfield{author}{\bibinfo{person}{Chuzhe Tang}, \bibinfo{person}{Youyun Wang}, \bibinfo{person}{Zhiyuan Dong}, \bibinfo{person}{Gansen Hu}, \bibinfo{person}{Zhaoguo Wang}, \bibinfo{person}{Minjie Wang}, {and} \bibinfo{person}{Haibo Chen}.} \bibinfo{year}{2020}\natexlab{}.
\newblock \showarticletitle{XIndex: a scalable learned index for multicore data storage}. In \bibinfo{booktitle}{\emph{Proceedings of the 25th ACM SIGPLAN symposium on principles and practice of parallel programming}}. \bibinfo{pages}{308--320}.
\newblock


\bibitem[Wang et~al\mbox{.}(2020)]%
        {wang2020sindex}
\bibfield{author}{\bibinfo{person}{Youyun Wang}, \bibinfo{person}{Chuzhe Tang}, \bibinfo{person}{Zhaoguo Wang}, {and} \bibinfo{person}{Haibo Chen}.} \bibinfo{year}{2020}\natexlab{}.
\newblock \showarticletitle{SIndex: a scalable learned index for string keys}. In \bibinfo{booktitle}{\emph{Proceedings of the 11th ACM SIGOPS Asia-Pacific Workshop on Systems}}. \bibinfo{pages}{17--24}.
\newblock


\bibitem[Wongkham et~al\mbox{.}(2022)]%
        {updatableindexready}
\bibfield{author}{\bibinfo{person}{Chaichon Wongkham}, \bibinfo{person}{Baotong Lu}, \bibinfo{person}{Chris Liu}, \bibinfo{person}{Zhicong Zhong}, \bibinfo{person}{Eric Lo}, {and} \bibinfo{person}{Tianzheng Wang}.} \bibinfo{year}{2022}\natexlab{}.
\newblock \showarticletitle{Are updatable learned indexes ready?}
\newblock \bibinfo{journal}{\emph{arXiv preprint arXiv:2207.02900}} (\bibinfo{year}{2022}).
\newblock


\bibitem[Wu et~al\mbox{.}(2021)]%
        {lipp}
\bibfield{author}{\bibinfo{person}{Jiacheng Wu}, \bibinfo{person}{Yong Zhang}, \bibinfo{person}{Shimin Chen}, \bibinfo{person}{Jin Wang}, \bibinfo{person}{Yu Chen}, {and} \bibinfo{person}{Chunxiao Xing}.} \bibinfo{year}{2021}\natexlab{}.
\newblock \showarticletitle{Updatable learned index with precise positions}.
\newblock  \bibinfo{volume}{14}, \bibinfo{number}{8} (\bibinfo{date}{April} \bibinfo{year}{2021}), \bibinfo{pages}{1276–1288}.
\newblock
\showISSN{2150-8097}
\urldef\tempurl%
\url{https://doi.org/10.14778/3457390.3457393}
\showDOI{\tempurl}


\bibitem[Wu et~al\mbox{.}(2022)]%
        {wu2022nfl}
\bibfield{author}{\bibinfo{person}{Shangyu Wu}, \bibinfo{person}{Yufei Cui}, \bibinfo{person}{Jinghuan Yu}, \bibinfo{person}{Xuan Sun}, \bibinfo{person}{Tei-Wei Kuo}, {and} \bibinfo{person}{Chun~Jason Xue}.} \bibinfo{year}{2022}\natexlab{}.
\newblock \showarticletitle{NFL: robust learned index via distribution transformation}.
\newblock \bibinfo{journal}{\emph{arXiv preprint arXiv:2205.11807}} (\bibinfo{year}{2022}).
\newblock


\end{thebibliography}

\end{document}